\documentclass[%
 reprint,
 superscriptaddress,
 nobibnotes,
 amsmath,amssymb,
 aps,
 prx
]{revtex4-2}

\usepackage{graphicx}
\usepackage{color}
\usepackage{dcolumn}
\usepackage{bm}
\usepackage{braket}
\usepackage{amsthm}
\usepackage{amssymb}
\usepackage{natbib}
\usepackage{siunitx}
\usepackage{booktabs}
\usepackage{mhchem}
\usepackage{threeparttable}
\usepackage{subfigure}
\usepackage{comment}
\usepackage[normalem]{ulem}

\theoremstyle{definition}
\newtheorem{definition}{Definition}[section]
\theoremstyle{plain}

\newtheorem{proposition}[definition]{Proposition}

\theoremstyle{definition}

\theoremstyle{remark}

\begin{document}

\preprint{APS/123-QED}

\title{A Gauss-Newton based Quantum Algorithm for Combinatorial Optimization}

\author{Mitsuharu Takeori}
\email{Mitsuharu.Takeori@ibm.com}
\affiliation{%
 IBM Quantum, 19-21 Nihonbashi Hakozaki-cho, Chuo-ku, Tokyo, 103-8510, Japan
}%
\author{Takahiro Yamamoto}%
\affiliation{%
 IBM Quantum, 19-21 Nihonbashi Hakozaki-cho, Chuo-ku, Tokyo, 103-8510, Japan
}%
\author{Ryutaro Ohira}%
\affiliation{%
 IBM Quantum, 19-21 Nihonbashi Hakozaki-cho, Chuo-ku, Tokyo, 103-8510, Japan
}%
\author{Shungo Miyabe}%
\affiliation{%
 IBM Quantum, 19-21 Nihonbashi Hakozaki-cho, Chuo-ku, Tokyo, 103-8510, Japan
}%

\date{March 25, 2022}

\begin{abstract}

In this work, we present a Gauss-Newton based quantum algorithm (GNQA) for combinatorial optimization problems that, under optimal conditions, rapidly converges towards one of the optimal solutions without being trapped in local minima or plateaus.
Quantum optimization algorithms have been explored for decades, but more recent investigations have been on variational quantum algorithms, which often suffer from the aforementioned problems.
Our approach mitigates those by employing a tensor product state that accurately represents the optimal solution, and an appropriate function for the Hamiltonian, containing all the combinations of binary variables.
Numerical experiments presented here demonstrate the effectiveness of our approach, and they show that GNQA outperforms other optimization methods in both convergence properties and accuracy for all problems considered here.
Finally, we briefly discuss the potential impact of the approach to other problems, including those in quantum chemistry and higher order binary optimization.

\end{abstract}


\maketitle


\section{Introduction}\label{Sec:introduction}

The combinatorial optimization models are versatile, and they play an important role in many industry relevant problems.
Therefore, various mathematical approaches have been proposed to build such models \cite{korte2011combinatorial, kochenberger2014unconstrained}.
Notably, one of the most effective analytical approaches is to formulate the problem as continuous optimization problems, a strategy that is still been actively investigated \cite{li2018combinatorial, goto2019combinatorial, PhysRevLett.122.040607, aramon2019physics}.

More recently, algorithms for optimization problems that seek to take advantage of quantum computers have been proposed.
Many of them are variational quantum algorithms (VQA) \cite{PhysRevA.92.042303, moll2018quantum, cerezo2021variational}, designed to solve the problem as a continuous optimization problem, similar to the above mentioned approach. Two widely studied VQA for combinatorial optimization problems are variational quantum eigensolver (VQE) \cite{peruzzo2014variational} and quantum approximate optimization algorithm (QAOA) \cite{farhi2014quantum}.
Notably, there have been recent efforts to improve upon the performance of these VQAs ~\cite{barkoutsos2020improving, amaro2022filtering}.
Although, we have not seen a clear evidence, supporting the advantage of these quantum algorithms over their classical counterparts, VQAs have paved the way for demonstrating the potential of real quantum processors through relevant scientific experiments.

VQAs have adopted many advantageous traits from continuous optimization problems, but there are two notable problems.
(1) VQAs deal with nonlinear continuous optimization problems, making them vulnerable to problems such as local minima and barren plateau \cite{mcclean2018barren, cerezo2021variational}.
The local minima decrease accuracy, and the barren plateau reduces computational efficiency, both of which are problematic.
(2) If the target Hamiltonian is close to a general real symmetric matrix with no symmetric structure, many circuit parameters are required.
In the worst case scenario, the number of circuit parameters needed will exceed the advantage gained by the use of quantum computers.

In this paper, we propose a Gauss-Newton based quantum algorithm (GNQA) for quadratic unconstrained binary optimization (QUBO), a particular class of comibinatorial optimization problem of great interest.
We show that this approach enables fast and stable convergence to one of the optimal solutions without being trapped in local minima or plateau.

To begin, we employ an ansatz parameterized by $N$ parameters, where $N$ is the size of the QUBO problem.
This means the dimension of the problem is only logarithm of the Hilbert space of the Hamiltonian.
Note that for QUBO problems, this ansatz can accurately represent one of the optimal solutions.

Next, we describe the advantage of our algorithm that is based on the Gauss-Newton method.
The ansatz that we use circumvents the problem of matrix inversion that limits the use of the method for large-scale problems.
In general, the Gauss-Newton method shows quadratic convergence in the neighborhood of the solution.
However, it requires the gradient and the inverse of the Gaussian matrix (or equivalently, the Fisher information), limiting its potential applicability in various industry relevant problems.
Remarkably, the inverse of the Gaussian matrix in our representation is just the identity matrix.
In addition, gradients are easily obtained using the parameter-shift rule \cite{li2017hybrid, mitarai2018quantum, schuld2019evaluating, mitarai2019methodology}.

Finally, we incorporate the advantage of the inverse power iteration algorithm, seen in eigenvalue algorithms, into the iterative process of the Gauss-Newton method. 
This method converges to the optimal solution without being trapped in local minima or plateaus in a small number of steps.
We note that the transformation of the Hamiltonian using nascent delta functions or hyper functions plays an important role, and they can be efficiently constructed on a quantum computer \cite{low2019hamiltonian, gilyen2019quantum, martyn2021grand}.

In this work, we investigated the potential of GNQA on various practical examples, such as the max cut problem, by measuring its performance against other quantum optimization algorithms.
The result presented here will show that our method outperforms other approaches in both convergence properties and accuracy for all problems considered here.
Finally, our approach also shows great promise in other important areas, including higher-order optimizations and quantum chemical calculations. 
We will briefly discuss such potentials.

Our paper is organized as follows.
We will briefly introduce quadratic unconstrained binary optimization problem in the next section.
In Sec.~\ref{Sec:classicalVQE}, we will show that our wavefunction is constructed as a tensor product state, and using such a state, we can propose efficient quantum-inspired algorithms that can be executed on classical computers.
The main result of this paper, the GNQA, is presented in Sec.~\ref{Sec:Main}.
Following the discussion on GNQA, we will show how the transformation of the Hamiltonian can be constructed in Sec.~\ref{Sec:construction-f(H)}.
Then, we will conclude with the presentation of numerical results in Sec.~\ref{Sec:main-results} and a short discussion and future perspectives in Sec.~\ref{Sec:discussion}.

\section{Preliminaries: Quadratic Unconstrained Binary Optimization Problem} \label{Sec:preliminaries}

We begin with a brief introduction of quadratic unconstrained binary optimization (QUBO) problem, which is expressed as
\begin{equation}
\label{qubo_orig}
    \begin{aligned}
        &\text{minimize} && x^T Q x \\
        &\text{subject to} && x_j \in \lbrace 0, 1 \rbrace &&&(j=0,\ldots,N-1),
    \end{aligned}
\end{equation}
where $Q = (q_{ij})_{i \leq j}$ is an $N\times N$ upper triangular matrix and $x = (x_j)_{0 \leq j < N}$ is a binary vector of $\mathbb{R}^N$.
To proceed, the problem is converted into the Ising model with the Hamiltonian defined as
\begin{equation}
    H = \sum_i h_i \sigma^z_i + \sum_{i<j}q_{i j} \sigma^z_i \sigma^z_j,
\end{equation}
where $\sigma^z_i$ is a Pauli matrix acting on the $i$-th qubit, and $h_i$ is defined as 
\begin{equation}
    h_i = - \sum_j \left( q_{i j} + q_{j i} \right).
\end{equation}
Here, $H$ is a linear operator acting on the $2^N$-dimensional Hilbert space $\mathcal{H} = \mathbb{R}^{2 \otimes N}$. 
Throughout this paper, we identify the Hilbert space $\mathcal{H}$ with $\mathbb{R}^M$, $M=2^N$.
In particular, the Hamiltonian $H$ is represented as a diagonal matrix by identifying $\mathcal{H} \simeq \mathbb{R}^M$.

Finally, we have the following optimization problem
\begin{equation} 
\label{variational_problem}
    \begin{aligned}
        &\text{minimize} && \braket{\xi|H|\xi} \\
        &\text{subject to} && \| \xi \| = 1, &&& \ket{\xi} \in \mathcal{H}.
    \end{aligned}
\end{equation}
which can be expressed as
\begin{align}
\label{final}
    \underset{x_k \in \lbrace 0, 1 \rbrace} {\text{min}} x^T Q x
    = \frac{1}{4} \left( -H_0 + \underset{\Vert \xi \Vert = 1} {\text{min}} \braket{\xi|H|\xi} \right).
\end{align}
Here $H_0$ is given by
\begin{align}
    H_0 = -2 \sum_i q_{ii} - \sum_{i<j} q_{ij}.
\end{align}

It is important to note that there is a correspondence between the left and the right solutions of Eq.~(\ref{final}).
Assuming no degeneracy, the right optimal solution can be represented as a one-hot vector
$\xi_{\mathrm{sol}} = (0,\ldots,0,1,0,\ldots,0)$ 
in $\mathcal{H}\simeq\mathbb{R}^M$, i.e., one of the orthonormal bases. 
Then, the index of the element with $\ket{\xi_{\mathrm{sol}}}$ of 1 corresponds to the decimal representation of the binary vector, the left solution.

\section{An Efficient Variational Algorithm}\label{Sec:classicalVQE}

The main result of this section is to show that our wavefuction is expressed as a tensor product state.
We will also show that quantum-inspired algorithms constructed using such a state, namely the gradient descent method, the modified Newton's method, and the natural gradient method, are powerful algorithms for combinatorial optimization problems, executed on classical computers.
However, they still suffer from problems associated with local minima or plateau.
Such problems will be addressed in the next section.

\subsection{A Representation of the Optimal Solution}

Here, we propose a variational approach to solve Eq.~(\ref{variational_problem}) using a state parameterized by $N$ circuit parameters, $\ket{\varphi(\theta)}$.
If the optimal solution is non-degenerate, $\ket{\varphi(\theta)}$ can be expressed as the following tensor product state,
\begin{align}
	\ket{\varphi(\theta)} = \ket{s(\theta_0)} \otimes \cdots \otimes \ket{s(\theta_{N-1})},
	\label{ansatz}
\end{align}
where $\ket{s(a)} = (\cos(a), \sin(a))$ and $\theta = (\theta_j)_{0 \leq j < N}$, $0 \leq \theta_j \leq \frac{\pi}{2}$. 
Otherwise, the solution can be expressed as a superposition of eigenstates $\ket{\xi_j}$ having the lowest eigenvalues,
\begin{align}
    \ket{\xi_{\mathrm{sol}}} = \sum_j \tau_j \ket{\xi_j}, 
\end{align}
where $\sum_j \vert \tau_j \vert^2 = 1$. 
Note that Eq.~(\ref{ansatz}) represents a pure state exactly, and thus, our approach works well for QUBO problems.

Given the above tensor product state, we now discuss the computational complexity of our algorithm.
The majority of the burden is placed on the calculation of the expectation value of the Hamiltonian $H$ with respect to $\ket{\varphi(\theta)}$, which can be expressed using harmonic functions in a following manner,
\begin{align}
	\braket{\varphi(\theta) |H| \varphi(\theta)}
	&=
	\sum_{i<j}q_{ij} \cos(2\theta_i) \cos(2\theta_j) \nonumber \\
	&\qquad -\sum_{i \leq j} q_{ij} \left( \cos(2\theta_i) + \cos(2\theta_j) \right).
	\label{reduction_formula}
\end{align}
Let $\mathcal{K} = [0, \pi/2]^N$ denote the parameter space. 
Then, the maximum principle states that the optimal solution can be found at one of the vertices of $\mathcal{K}$, 
where the above tensor product state becomes an orthonormal basis for $\mathbb{R}^N$. 
Clearly, the computational cost of Eq.~(\ref{reduction_formula}) is $O(N^2)$, not $O(M)$, where $M$ is the dimension of the Hilbert space.
Furthermore, if the matrix $Q$ of the QUBO problem is sparse with $\kappa$ non-zero elements, it is possible to compute the expectation values with $O(\kappa)$ complexity.

Now, the optimization problem of Eq.~(\ref{variational_problem}) can be reformulated with a parametrized state to give
\begin{equation}
    \label{optimization_problem_form2}
    \begin{aligned}
        &\text{minimize} && x^T A x + b \cdot x \\
        &\text{subject to} && x_k = \frac{1}{2} \left(1 - \cos(2 \theta_k) \right),
    \end{aligned}
\end{equation}
where $A$ is a matrix derived from the off-diagonal elements of $Q$ and $b$ is a vector of diagonal elements of $Q$.
We expressed the solution of Eq.~(\ref{variational_problem}) with the optimized parameter $\theta^*$ as $\ket{\xi^*} = \ket{\varphi(\theta^*)}$.
Finally, using $\theta^*$ we also obtain the solution of Eq.~(\ref{optimization_problem_form2}),
\begin{align}
    x^* = \frac{1}{2} \left(1 - \cos(2 \theta^*) \right),
\end{align}
which is a binary vector.
Note that the function on the right hand side is  a component-wise function.

\subsection{Gradient Descent Method}\label{Subsec:gradient-descent}
The variational ansatz (Eq.~(\ref{ansatz})) allows us to reformulate Eq.~(\ref{variational_problem}) as a continuous optimization problem,
\begin{align}
	\min_{\theta \in \mathcal{K}} L(\theta),
\end{align}
where $L(\theta) = \braket{\varphi(\theta) |H| \varphi(\theta)}$.
Then, the gradient descent method for the objective function $L$ is given by
\begin{align}
    \theta^{(n+1)} = \theta^{(n)} - \eta \, \nabla L ( \theta^{(n)} ),
\end{align}
where $\eta$ is a step size. 
Note that in many variational methods, the initial state is typically set to $\varphi_\mathrm{ini} = \varphi(\theta^{(0)})$, where $\theta^{(0)} = (\pi/4, \cdots, \pi/4)$.

It follows from Eq.~(\ref{reduction_formula}) that the computational complexity in obtaining the gradient of $L$ is also $O(N^2)$, making the gradient decent approach tractable on a classical computer.
In fact, if we set $y = \nabla L(\theta)$, the $j$-th element of $y$ is given by
\begin{equation} 
    y_j 
    = 2 \sin(2 \theta_j) \left( 2q_{j j} + \sum_{k=1}^{N-1} \tilde{q}_{j k} (1- \cos(2 \theta_k)) \right),
\end{equation}
where
\begin{equation} 
    \tilde{q}_{j k} =
	\begin{cases}
        q_{j k} & j \leq k  \\
        q_{k j} & j > k
    \end{cases}.
\end{equation}
Moreover, if the matrix $Q$ is sparse with $\kappa$ non-zero elements, the complexity can be reduced even further to $O(\kappa)$ by using the following update formula for $(j,k)$, where $q_{j,k} \neq 0$ and $y$ is initially set to $0$:
%
\begin{align}
    y_j &\leftarrow y_j +
	\begin{cases}
        2 q_{j k} \sin(2 \theta_j) & j = k  \\
        2 q_{j k} \sin(2 \theta_j) (1 - \cos(2 \theta_k)) & j > k
    \end{cases}, \\
    y_k &\leftarrow y_k +
	\begin{cases}
        2 q_{j k} \sin(2 \theta_k) & j = k  \\
        2 q_{j k} \sin(2 \theta_k) (1 - \cos(2 \theta_j)) & j > k
    \end{cases}.
\end{align}

Although this algorithm can be efficiently executed on a classical computer, as demonstrated above, we are still faced with the following challenges.
(1) A large number of iterations are required to find an accurate solution. This problem becomes more pronounced with the presence of barren plateau \cite{mcclean2018barren, cerezo2021variational}.
(2) Presence of local minima or stationary points can effect the accuracy.

Such problems have been extensively studied, and various algorithms have been proposed to overcome the challenges.
For example, acceleration methods, such as the standard momentum method, Nesterov's Accelerated method \cite{nesterov2003introductory, NEURIPS2018_44968aec}, or ADAM \cite{kingma2014adam}, help mitigate problems associated with barren plateau.
Quasi-Newton methods such as the BFGS \cite{liu1989limited, li2001modified} are also used for the same purpose.
To mitigate the problems associated with local minima, the Basin-Hopping algorithm \cite{wales1997global}, which incorporates local minimization steps within the global optimization in a stochastic manner, has enjoyed great success.
Moreover, the Bayesian optimization \cite{shahriari2015taking} is another similar global optimization technique.

\begin{figure}[t]
\centering
    \includegraphics[width=8.5cm]{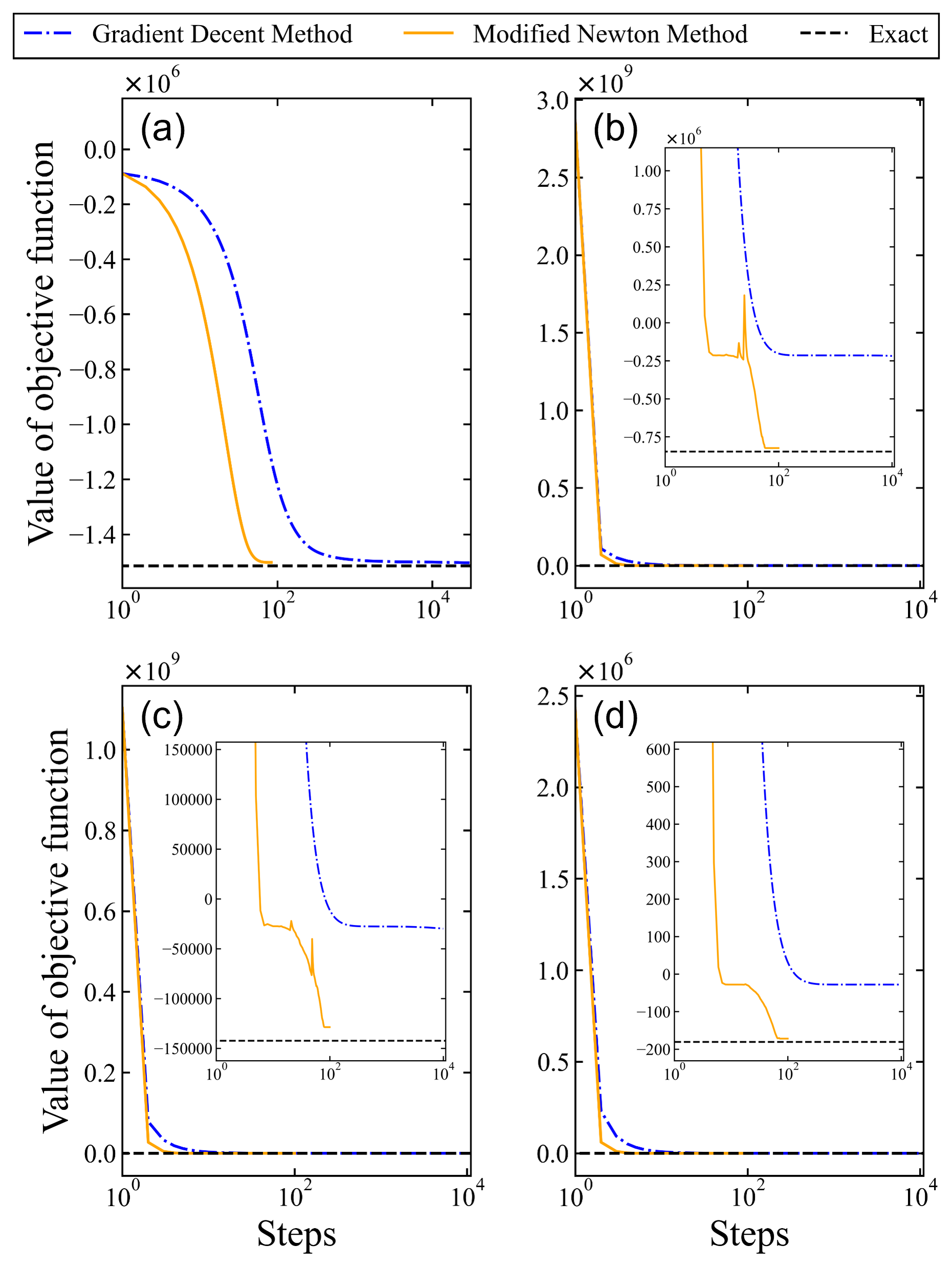}
    \caption{\label{Fig:classical_exp_1} 
    We plot the convergence behavior of the gradient descent method (blue dashed dotted line) and the modified Newton's method (orange solid line) for different QUBO problems of varying size $N$.
    The problems employed were (a) a benchmark QUBO problems and three other QUBO problems including (b) a quadratic assignment problem ($N=10000$), (c) a traveling salesman problem ($N=22201$), and (d) a N-queens problem ($N=32761$).
    }
\end{figure}

\subsection{Modified Newton's method}\label{Subsec:modified-newton}

We now describe an effective method that employs the information of both the gradient and its Jacobian, namely the Hessian of $L$, to optimize the convergence behavior of our algorithm.
Using the state given by Eq.~(\ref{ansatz}), we can efficiently apply the Newton's method and find a shortest path to a point on the Riemannian manifold where $\nabla L(\theta)=0$ (see Appendix~\ref{Appendix:Geo-Newton}).
However, it is important to note that this solution is not necessarily the optimal solution.

Another advantage of our formulation is that the Hessian of the objective function $L$ can be computed efficiently.
By setting $Y$ as the Hessian of $L$, 
we arrive at the update rule based on the Newton's method
\begin{align}
    \theta^{(n+1)} = \theta^{(n)} - Y(\theta^{(n)})^{-1} \nabla L(\theta^{(n)}),
\end{align}
which has the following formulae for computing the elements of $Y$,
\begin{align}
    Y_{j j}(\theta) 
    &= 4 \cos(2 \theta_j) 
       \left( 2q_{j j} + \sum_{k=1}^{N-1} q_{j k} (1- \cos(2 \theta_k)) \right), \\
    Y_{j k}(\theta) 
    &= 4 q_{j k} \sin(2 \theta_j) \sin(2 \theta_k), \quad j \neq k.
\end{align}
It is important to note that this algorithm shows quadratic convergence in the neighbourhood of the solution \cite{Kantorovich, traub1979convergence}.
However, the radius of convergence is too small to guarantee the global convergence property, necessitating the combination of line search algorithms. 

To circumvent the problem, particularly for the state in Eq.~(\ref{ansatz}), we replace the Hessian matrix $Y$ with
\begin{align}
    \hat{Y} = Y + \nu \, \| \nabla L \| I,
\end{align}
where $I$ denotes the identity operator and $\nu > 0$ is a real value.
This improves the overall convergence property and stability of the method without sacrificing its performance in the neighborhood of the solution.
Note that this benefit is provided by the Levenberg–Marquardt algorithm \cite{levenberg1944method, marquardt1963algorithm} that bridges the Newton's algorithm and the gradient descent method.

We now discuss the computational complexity of this modified Newton's method.
Since the computational complexity of directly obtaining $\hat{Y}^{-1}$ is $O(N^3)$, we employ an iterative approach that uses matrix-vector multiplication.
If $Q$ is sparse with $\kappa$ non-zero elements, we can compute $Y \times v$ with a computational complexity of $O(\kappa)$, where $v$ is a vector.
This is accomplished by using the following update formula for $(j,k)$ where $q_{j k} \neq 0$ and $b$ is initially set to $0$:
\begin{align}
    b_j &\leftarrow b_j +
	\begin{cases}
        4 q_{j k} \cos(2 \theta_j) v_j & j = k  \\
        4 q_{j k} \cos(2 \theta_j) (1 - \cos(2 \theta_k)) v_j \\
        \qquad + 4 q_{j k} \sin(2 \theta_j) \sin(2 \theta_k) v_k & j > k
    \end{cases} \\
    b_k &\leftarrow b_k +
	\begin{cases}
        4 q_{j k} \cos(2 \theta_k) v_k & j = k  \\
        4 q_{j k} \cos(2 \theta_k) (1 - \cos(2 \theta_j)) v_k \\ 
        \qquad + 4 q_{j k} \sin(2 \theta_k) \sin(2 \theta_j) v_j & j > k
    \end{cases}
\end{align}
Then, we can employ iterative methods such as MINRES  \cite{paige1975solution, simoncini2007recent} to get $\hat{Y}^{-1} v$.
Finally, we use the inverse of the diagonal matrix consisting of $\mathrm{diag}(\hat{Y})$ as the pre-conditioner for MINRES.
This is valid since the off-diagonal elements of $\hat{Y}$ vanishes as $\theta$ approaches the solution.
In many cases, we observed that such a pre-conditioner considerably improves the rate of convergence, especially in the neighborhood of solution.

There are two main advantages to this approach; (1) the method is efficient even executed on classical systems, and (2) it converges faster than the gradient descent method.
However, the method still has problems associated with local minima.

To demonstrate the efficiency of the modified Newton's method, its performance was compared against the gradient descent method.
The result is shown in Fig.~\ref{Fig:classical_exp_1}.
To obtain these results we employed (a) a benchmark QUBO problems of size $N=2500$ found in OR-library \cite{beasley1990or} and three other QUBO problems \footnote{Problems available at https://qubo.cs.hiroshima-u.ac.jp/qubo-examples} that include (b) a quadratic assignment problem ($N=10000$), (c) a traveling salesman problem ($N=22201$), and (d) a N-queens problem ($N=32761$).
For all examples, the modified Newton's approach exhibits faster convergence and better accuracy compared with the gradient descent method.
Note that in these calculations $\theta$ was initially set to  $(\pi/4, \cdots, \pi/4)$, and with such a condition, the algorithm was not able to find the optimal solution.
However, one could improve the result of these calculations by using a better initial point or incorporating global optimization techniques like  the ones discussed in Refs.~\cite{wales1997global, bergstra2011algorithms, hansen2016cma}.
They are implemented in the libraries SciPy and Optuna \cite{akiba2019optuna}.

\subsection{Generalized natural gradient method for eigenvalue problems}\label{Subsec:natural-gradient}

The VQE solves the optimization problem by minimizing the Rayleigh quotient as formulated in Eq.~(\ref{variational_problem}).
Alternatively, we can obtain the ground state by solving the following non-linear equation:
\begin{align} \label{geo_eq}
    (H - E_0) \ket{\xi} = 0,
\end{align}
where $E_0 = \lambda_0 I$ is the ground state energy of $H$.
Now we will formulate Eq.~(\ref{geo_eq}) as an optimization problem, and we begin with the following eigenvalue problem:
\begin{equation} \label{eigenval_prob}
    \begin{aligned}
        &\text{minimize} && \braket{\xi| (H - E_0) |\xi} \\
        &\text{subject to} && \Vert \xi \Vert = 1, &&& \xi \in \mathcal{H}.
    \end{aligned}
\end{equation}
%
In our approach, we employ $\ket{\varphi(\theta)}$ defined by Eq.~(\ref{ansatz}) to obtain the following objective function
\begin{align}
    \hat{L}(\theta) = \braket{\varphi(\theta) | (H - E_0) | \varphi(\theta)}.
\end{align}
Then, we set up the equation $F = 0$ by introducing
\begin{align}
    F(\theta) 
    = \dfrac{\partial \hat{L}(\theta)}{\partial \varphi(\theta)}
    = 2 (H - E_0) \ket{\varphi(\theta)}.
\end{align}
Additionally, we define the positive definite symmetric matrix
\begin{align}
    \begin{split}
    G 
    &= J_{\varphi}^T J_F = 2 J_{\varphi}^T (H - E_0) J_{\varphi} \\
    &= 2 \left( J_{\varphi}^T H J_{\varphi} - E_0 \right),
    \end{split}
\end{align}
where $J_\varphi$ and $J_F$ denote the Jacobi matrices in terms of $\varphi$ and $F$, respectively, namely,
\begin{equation}
    J_{\varphi}(\theta) = \dfrac{\partial}{\partial \theta} \ket{\varphi(\theta)},
    \quad
    J_F(\theta) = \frac{\partial}{\partial \theta} F(\theta).
\end{equation}
Note that here we use the relation $J_\varphi^T J_\varphi = I$, which can be derived from a straightforward calculation.

Using $(\mathcal{K}, G)$ as a Riemannian manifold, 
we can define the gradient descent method on this curved space by
\begin{align}
    \theta^{(n+1)} 
    = 
    \theta^{(n)} - \eta \, \nabla_G \hat{L}(\theta^{(n)}),
\end{align}
where
\begin{align}
    \nabla_G = G^{-1} \nabla.
\end{align}
Under the condition that the distance, given in terms of the Riemannian metric $G$,
\begin{equation}
    \| \theta \|_G = \theta^T G \theta .
\end{equation}
is constant, the direction which maximizes the change of the objective function $\hat{L}$ is $\nabla_G \hat{L}$.
Note that this approach is equivalent to the natural gradient method \cite{amari1998natural, insight_natural_grad, stokes2020quantum} that is often used in the field of machine learning and deep learning with the least squares method.

One of the advantages of natural gradient method is its property to avoid plateaus. 
Moreover, when $\eta=1$, this method can be viewed as a way to formulate the eigenvalue problem given above as a Gauss-Newton method.
Notably, the method shows a quadratic convergence behavior in the neighborhood of the solution if the optimal solution satisfies the condition that $F=0$.
The convergence behavior of the Gauss-Newton method has been extensively studied ~\cite{chen2005convergence, alvarez2008unifying}.

It is worthwhile to note the difference between the Newton's method discussed in Sec.~\ref{Subsec:modified-newton} and the natural gradient method of this section.
The Newton's method identifies the closest region, satisfying $\nabla L = 0$, from the initial starting point, and it gives the shortest path to that region. 
On the other hand, the natural gradient approach based on Eq.~(\ref{eigenval_prob}) identifies the point satisfying $F=0$, and it provides the optimal path to that point.
This difference will be highlighted in later examples (see Fig.~\ref{fig:comparison_algorithms} in Sec.~\ref{Sec:main-results}). 

The computational cost of our method depends on the problem size $N$, which is equivalent to the modified Newton's method (Sec.~\ref{Subsec:modified-newton}). 
$\braket{\varphi | \varphi} = 1$ implies $J^T_{\varphi} \ket{\varphi} = 0$, and therefore, we have
\begin{align}
    \nabla \hat{L}(\theta)
    = \nabla L(\theta) - \lambda_0 J^T_{\varphi}(\theta) \ket{\varphi(\theta)} 
    = \nabla L(\theta).
\end{align}
Clearly, the computational cost to obtain the gradient of $\hat{L}$ and $L$ is equivalent.
Now, if we define $\tilde{G}$ as $\tilde{G} = J_{\varphi}^T H J_{\varphi}$, it can be explicitly computed using the following formula:
\begin{align}
    \tilde{G}_{k k}(\theta) &= 
	    \sum_{i<j} q_{i j} C_{k i} C_{k j} 
	    - \sum_{i \leq j} q_{i j} \left( C_{k i} + C_{k j} \right)\\
    \tilde{G}_{k, l}(\theta) &= 
        q_{k l} \sin(2 \theta_{k}) \sin(2 \theta_{l})
        \quad (k \neq l),
\end{align}
where
\begin{align}
	    C_{k j} = 
	    \begin{cases}
            \cos(\pi - \theta_j) & j=k  \\
            \cos(\theta_j) & j \neq k
        \end{cases}. 
\end{align}
$G$ is related to the Hessian $Y$, discussed in the previous section, by 
\begin{align} \label{Gauss-Hess_relation}
    G = Y_0 + \frac{1}{2} Y_1 + 2(L - \lambda_0), 
\end{align}
where $Y_0$ and $Y_1$ are matrices consisting of diagonal and off-diagonal elements of $Y$, respectively.
Thus, the computational complexity in obtaining $\hat{G}$ is $O(N^2)$, or $O(\kappa)$ in the case of sparse $Q$ with $\kappa$ non-zero elements.
As discussed previously, the iterative methods are more advantageous in computing the inverse of $G$ because the computational cost of the matrix-vector multiplication in terms of $G$ is essentially the same as that of $Y$ due to Eq.~(\ref{Gauss-Hess_relation}).

The disadvantage of the method is that the information about the minimum eigenvalue $E_0$ must be known priori.
While there is an accurate way to compute $E_0$ through exponentiation of the Hamiltonian, here we introduce an alternative statistical way to estimate $E_0$ on a classical system.
First, notice that Eq.~(\ref{reduction_formula}) implies that the Hamiltonian $H$ possesses the following properties:
\begin{gather}
	\mathrm{Tr}(H) = 0, \label{H_sum} \\
	\left\| H \right\|_F = M^{\frac{1}{2}} \sqrt{ \sum_{i<j} q_{ij}^2 + \sum_i h_i^2 },
	\label{H_norm}
\end{gather}
where $\Vert \cdot \Vert_F$ is the Frobenius norm. 
Since the Hamiltonian $H$ is represented as a diagonal matrix on $\mathcal{H}\simeq\mathbb{R}^M$, Eq.~(\ref{H_sum}) implies that the mean of $d_H$ is $0$, where $d_H$ is a vector consisting of $M$ diagonal elements of $H$. 
Therefore, Eq.~(\ref{H_norm}) gives the variance or standard deviation $\sigma$ of $d_H$, i.e.
\begin{gather}
	E(d_H) = 0,  \label{H_average}\\
	V(d_H) = \sum_{i<j} q_{ij}^2 + \sum_i h_i^2, \label{H_std}
\end{gather}
The above equations imply that the computational complexity for obtaining the variance of $d_H$ is $O\left( N^2 \right)$.
Note that the approach of this subsection requires one to estimate the range of $d_H$, and guess the minimum energy.
In many cases, this range is typically $3 \sim 10 \sigma$.
Finally, if there is a large discrepancy between the approximated value and the true minimum, the convergence rate slows down and the possibility of getting stuck in a local solution increases.

The generalized natural gradient method discussed here has the following advantages.
(1) It can efficiently be executed on a classical computer.
(2) In general, the convergence is faster than the gradient descent method. 
In particular, when the exact value of $E_0$ is known, the convergence is quadratic in the neighborhood of the solution.
(3) It circumvents a plateau.
However, there are two main disadvantages.
(1) The minimum energy eigenvalue must be approximated with a sufficient enough accuracy, but it is difficult to obtain such a value using classical means.
(2) The problems associated with local minima persist.

\section{Gauss-Newton based quantum algorithm}\label{Sec:Main}

In the previous section, we presented quantum-inspired algorithms, built on the state defined by Eq.~(\ref{ansatz}), that can efficiently be executed on a classical computer.
We also described the disadvantages of such algorithms, namely the local minima and plateau.

In this section, we present an effective quantum algorithm for optimization problems, and this is the main result of this paper.
We will show that our algorithm converges rapidly to an optimal solution without being trapped in local minima or plateaus.
We begin this section with the description of the Gauss-Newton method.
Then, we will introduce an iterative approach that uses an approximation to the ground state created through a transformation of $H$.
We will show that this transformation can more efficiently be carried out on quantum computers, and it is one of the key advantages offered by this algorithm.

\subsection{Gauss-Newton Method} \label{Subsec:sec-converge}

The most effective way to obtain the solution of the QUBO problem Eq.~(\ref{variational_problem}), or other equivalent optimization problems, is to solve the following problem:
\begin{align} \label{modified_optimization_problem}
	\min_{\theta \in \mathcal{K}} \frac{1}{2} \left\Vert \ket{\varphi(\theta)} - \ket{\xi^*} \right\Vert^2,
\end{align}
where $\ket{\xi^*}$ is the ground state of $H$. 
Note that this requires the prior knowledge of the ground state, and thus, it is not practical.
However, the idea of estimating the optimal solution in terms of parameters is the general framework we use in our approach.

We now describe the Gauss-Newton method for this least squares problem.
Since $G = J_{\varphi}^T J_{\varphi} = I$ for the state defined by Eq.~(\ref{ansatz}), the metric for the corresponding Riemannian space is the same as that for the Euclidean space. 
Thus, we can consider the recurrence relation of the form,
\begin{align}
    \theta^{(n+1)} &=
    \theta^{(n)} - \eta \, J_{\varphi}^T(\theta^{(n)}) \left( \ket{\varphi(\theta^{(n)})} - \ket{\xi^*} \right) 
    \nonumber \\
    &= \theta^{(n)} + \eta \, J_{\varphi}^T(\theta^{(n)}) \ket{\xi^*}.
    \label{ideal_GN_method}
\end{align}
Note that this is the Guass-Newton iteration for $\eta=1$.
If we set $\theta^{(0)}$ as $(\pi/4,\ldots,\pi/4)$, the trajectory given by the above equation is a straight line to the optimal solution (see Fig.~\ref{fig:comparison_algorithms} in Sec.\ref{Sec:main-results}). 
To see this, let's suppose that the optimal solution is $\ket{\xi^*} = (1, 0, \cdots, 0)$ and $\theta^* = (0, \cdots, 0)$. 
Then, the $k$-th element of $y = J_{\varphi}^T(\theta) \ket{\xi^*}$ in terms of $\theta = (\omega, \cdots, \omega)$ is given by
\begin{align} \label{convergence_along_line_0}
    y_k 
    &= -\cos(\theta_0) \cdots \sin(\theta_k) \cdots \cos(\theta_{N-1}) \nonumber \\
    &= - \cos(\omega)^{N-1} \sin(\omega).
\end{align}
This implies that the update values for all elements of $\theta^{(n)}$ at the $n$-th step are the same.
Moreover, we can see that the limit of the recurrence relation for any positive integer $\alpha > 0$,
\begin{align} \label{convergence_along_line}
    x^{(n+1)} = x^{(n)} - \eta \cos(x^{(n)})^{\alpha} \sin(x^{(n)}),
\end{align}
will go to zero as $n \rightarrow \infty$ for a point initially at $x^{(0)} = \pi/4$.
Notably, it shows a rapid convergence in the neighborhood of zero because $\cos(x_n) \approx 1$.

In summary, if the initial point of a trajectory was chosen to be at the center of the cube $\mathcal{K}$, the Gauss-Newton iteration will guide the trajectory, in a straight line, to one of the vertices of the cube where the parameter represents the optimal solution, exactly.
Moreover, $y$, used for updating $\theta$, does not vanish along this path except at the location of the optimal solution.

In general, the trajectory defined by Eq.~(\ref{ideal_GN_method}) does not necessarily lie on such a straight path.
To demonstrate the convergence behavior of the Gauss-Newton method in a more general situation, we discuss its property in the neighborhood of the solution when $\eta = 1$.
Let $\theta^*$ be the optimal solution of the above optimization problem, i.e. $\ket{\varphi(\theta^*)}=\ket{\xi^*}$. 
Then, we choose $\frac{1}{2} < C < 1$ such that
\begin{align}
    \left\| J_{\varphi}(\theta^{(n)}) - J_{\varphi}(\theta^*) \right\|
    <
    \frac{2C}{\sqrt{N}}
    \left\| J_{\varphi}(\theta^{(n)}) - J_{\varphi}(\theta^*) \right\|_F.
\end{align}
%
Next, we express the difference between the current point $\theta$ of the optimization and the optimal solution $\theta^*$ as
\begin{align}
    \begin{split}
    &\left\| J_{\varphi}(\theta) - J_{\varphi}(\theta_s) \right\|_F^2 \\
    &\quad = \mathrm{Tr}\left( 
        (J_{\varphi}(\theta) - J_{\varphi}(\theta_s))^T
        (J_{\varphi}(\theta) - J_{\varphi}(\theta_s)) 
        \right)\\
    &\quad = 2N \left( 1 - \prod_{k=0}^{N-1} \cos \left( (1-s)(\theta_k - \theta^*_k) \right) \right) \\
    &\quad \leq N \| (1-s) (\theta - \theta^*) \|^2,
    \end{split}
\end{align}
where $\theta_{s} = (1-s)\theta^* + s \theta$ ($0 \leq s \leq 1$). 
Note that the last inequality follows from the fact that the function $g \colon \mathbb{R}^N \rightarrow \mathbb{R}$, defined by
\begin{align}
    g(x) = \sum_k x_k^2 - 2 \left( 1 - \prod_k \cos(x_k) \right),
\end{align}
is a non-negative function that is $0$ if and only if $x=0$.
Now, we have
\begin{align}
    \begin{split}
    \left\| J_{\varphi}(\theta) - J_{\varphi}(\theta_s) \right\| 
    &< \frac{2C}{\sqrt{N}} \left\| J_{\varphi}(\theta) - J_{\varphi}(\theta_s) \right\|_F \\
    &\leq   2C (1-s) \| \theta - \theta^* \|.
    \end{split}
\end{align}
Since $J^T_{\varphi} J_{\varphi} = I$, we get $\| J_{\varphi} \| = 1$.
Therefore, we have
\begin{align}
    \begin{split}
    &\theta^{(n+1)} - \theta^* \\
    &\quad =
    \theta^{(n)} - \theta_{*} - J^T_{\varphi}(\theta^{(n)})
    \left( \ket{\varphi(\theta^{(n)})} - \ket{\varphi(\theta^*)} \right) \\
    &\quad =
    J^T_{\varphi}(\theta^{(n)}) \Big[
    J_{\varphi}(\theta^{(n)}) (\theta^{(n)} - \theta^*) +
    \ket{\varphi(\theta^{(n)})} - \ket{\varphi(\theta^*)}
    \Big] \\
    &\quad = 
    J^T_{\varphi}(\theta^{(n)}) \int_{0}^{1} \Big[
    J_{\varphi}(\theta^{(n)}) - J_{\varphi}(\theta^{(n)}_s)
    \Big] (\theta^{(n)} - \theta^*) \, ds,
    \end{split}
\end{align}
and
\begin{align}
    \begin{split}
    \left\| \theta^{(n+1)} - \theta^* \right\|
    &\leq
    \left\| \theta^{(n)} - \theta^* \right\|
    \int_0^1
    \left\| J_{\varphi}(\theta^{(n)}) - J_{\varphi}(\theta^{(n)}_s) \right\|
    \, ds \\ 
    &<
    C \left\| \theta^{(n)} - \theta^* \right\|^2,
    \end{split}
\end{align}
demonstrating the quadratic convergence of our approach.

\subsection{Approximate Ground State with Transformation of H}\label{Subsec:basic-form}

We now go back to Eq.~(\ref{modified_optimization_problem}), and we consider an iterative approach that uses an approximate ground state that is updated at every iteration step instead of using the true ground state.
Let $\ket{\xi_k}$ be the eigenvector of the Hamiltonian $H$ for $k=0, \ldots, M-1$ with the corresponding eigenvalue $\lambda_k$ such that $\lambda_0 < \lambda_1 < \cdots < \lambda_{M-1}$.
Then, an arbitrary state in $\mathcal{H}$ can be expressed as
\begin{align}
    \ket{\varphi} = \sum_{k=0}^{M-1} a_k \ket{\xi_k},
\end{align}
where $\sum_{k=0}^{M-1} a_k^2 = 1$.
Note that since the Hilbert space $\mathcal{H}$ is real, every coefficient $a_k$ is real.

Now we consider a transformation of $H$ using an analytical function $f$, denoted $f(H)$.
Next we define a nonlinear transformation $R_f$ associated with $f$ that maps one state to another through
\begin{align}
    R_f \left( \ket{\varphi} \right) = 
    \frac{f(H) \ket{\varphi}}{\left\| f(H) \ket{\varphi} \right\|}= 
    \sum_{k=0}^{M-1} c_k \ket{\xi_k},
\end{align}
where
\begin{align}
    c_k = \frac{a_k f(\lambda_k)}{\sqrt{\sum_j a_j^2 f(\lambda_j)^2}}.
\end{align}
Ideally, to obtain $R_f \left( \ket{\varphi} \right) = \ket{\xi^*}$, we would take a compactly supported continuous function for $f$ whose support is contained only in a neighborhood of $\lambda_0$, such as the Dirac delta function $\delta(x - \lambda_0)$.
However, in practice, we utilize functions that possess a weaker property,
\begin{align} \label{prop_f}
    \vert f(\lambda_0) \vert \gg  \vert f(\lambda_k) \vert
    \qquad k=1,\cdots,M-1,
\end{align}
and demonstrate its validity through experiments and analysis.

Now, the algorithm for the optimization problem defined in Eq.~(\ref{modified_optimization_problem}) reads
\begin{equation}
    \begin{aligned}
    \theta^{(n+1)}
    &=
    \theta^{(n)} - \eta \, G(\theta^{(n)})^{-1} 
    J^T_{\varphi}(\theta^{(n)}) \left( \ket{\varphi(\theta^{(n)})} - \ket{\zeta^{(n)}} \right), \\
    \ket{\zeta^{(n)}} 
    &= 
    R_f \left( \ket{\varphi(\theta^{(n)})} \right), \\
    G &= J_\varphi^{T} J_\varphi,
    \end{aligned}
\end{equation}
where $G = I$ in our case.
Thus, the algorithm based on the Gauss-Newton method is consequently formulated as follows:
\begin{align}
    \theta^{(n+1)}
    = \theta^{(n)} + J^T_{\varphi}(\theta^{(n)}) R_f ( \ket{\varphi(\theta^{(n)})} ).
\end{align}
Note that $\ket{\xi^*}$ is a fixed point in $R_f$.
Therefore, this algorithm is expected to exhibit behavior analogous to Eq.~(\ref{ideal_GN_method}) in the neighborhood of the optimal solution.
In the following subsections, we will introduce properties that merit the use of above approach.

\subsection{Error Analysis}\label{Subsec:err-estimation}

In this section we show that there is an error associated with the states created using $R_f$.
However, this error approaches zero as $\theta^{(n}$ approaches the optimal solution $\theta^*$ with increased iteration.

First we show the error associated with the states created using $R_f$.
We begin by defining a residual of $f(H)$ as
\begin{align}
    r = \sum_{k>1} \left( \frac{f(\lambda_k)}{f(\lambda_0)} \right)^2.
\end{align}
Then, we have
\begin{align}
    \| R_f ( \ket{\varphi(\theta)} ) - \ket{\xi^*} \|^2
    &= 2 \Big( 1 - \braket{\xi^* | R_f ( \ket{\varphi(\theta)} )} \Big) \nonumber \\
    &= 2 \left( 1 - C_f \, \varphi_0(\theta) \right)
\end{align}
with
\begin{align}
    C_f
    = \frac{f(\lambda_0)}{\sqrt{\sum_{k} f(\lambda_k)^2 \varphi_k(\theta)^2}}.
\end{align}
If $\varphi_0(\theta) \geq \varphi_k(\theta)$ for $k=1, \ldots, M-1$, we have
\begin{align*}
    \left( C_f \, \varphi_0(\theta) \right)^{-2}
    &= 1 + \sum_{k>1} 
    \left( \frac{f(\lambda_k)}{f(\lambda_0)} \right)^2 
    \left( \frac{\varphi_k(\theta)}{\varphi_0(\theta)} \right)^2 \nonumber \\
    &\leq 1 + r,
\end{align*}
so that, 
\begin{align*}
    \| R_f ( \ket{\varphi(\theta)} ) - \ket{\xi^*} \|^2
    \leq 2 - \frac{2}{\sqrt{1 + r}}.
\end{align*}
If $f(\lambda_0) \gg f(\lambda_k)$, then $r \approx 0$. 
Therefore, the state created using $R_f$ approaches the true ground state as $f(\lambda_0)$ becomes dominant.

Now, we explicitly show how this discrepancy between the true ground state and the state created via $R_f$ manifests in the convergence behavior of the algorithm.
We assume that the following condition holds for an infinitesimal number $\varepsilon$. 
\begin{align}
    \| R_f ( \ket{\varphi(\theta)} ) - \ket{\xi^*} \| < \varepsilon,
\end{align}
Then, since
\begin{multline}
    \theta^{(n+1)} - \theta^*
    =
    \theta^{(n)} - \theta^* \\
    - J^T_{\varphi}(\theta^{(n)}) \left( \ket{\varphi(\theta^{(n)})} - \ket{\varphi(\theta^*)} \right) \\
    + J^T_{\varphi}(\theta^{(n)}) \left( \ket{\zeta^{(n)}} - \ket{\xi^*} \right),
\end{multline}
we can show, as in Sec.~\ref{Subsec:basic-form}, that
\begin{align}
    \left\| \theta^{(n+1)} - \theta^* \right\|
    &<
    C \left\| \theta^{(n)} - \theta^* \right\|^2 + \varepsilon.
\end{align}
Therefore, there is an error $\varepsilon$ that appears at an iteration.

Now, we show this error $\varepsilon$, describing the difference between $R_f$ and the true ground state, decreases as $\theta^{(n)}$ approaches $\theta^*$. 
Therefore, in the vicinity of the solution, the algorithm is expected to have the same performance as Eq.(\ref{ideal_GN_method}).
More concretely, we show that $\theta \mapsto R_f ( \ket{\varphi(\theta)} )$ is a contraction mapping in the sense that there exists $0<q<1$ for which the following condition holds.
\begin{align} \label{Contraction mapping}
    \| R_f ( \ket{\varphi(\theta)} ) - R_f ( \ket{\varphi(\theta^*)} ) \|
    < q \| \theta - \theta^* \|.
\end{align}
To proceed, we note that
\begin{align*}
    C_f \geq
    \frac{f(\lambda_0)}{\sqrt{\sum_{k} f(\lambda_0)^2 \varphi_k(\theta)^2}} = 1
\end{align*}
holds, where the equality is valid if and only if $\theta=\theta^*$.
Then, we have
\begin{align}
    \begin{split}
    \| R_f ( \ket{\varphi(\theta)} ) - \ket{\xi^*} \|^2
    &= 2 \left( 1 - C_f \, \varphi_0(\theta) \right) \\
    &\leq \left( 1 - \varphi_0(\theta) \right) \\
    &= \| \ket{\varphi(\theta)} - \ket{\xi^*} \|^2 \\
    &= \| \ket{\varphi(\theta)} - \varphi(\theta^*) \|^2,
    \end{split}
\end{align}
On the other hand, we can also have 
\begin{align}
    \begin{split}
    \| \ket{\varphi(\theta)} - \varphi(\theta^*) \|^2
    &= 2 \left( 1 - \prod_{k=0}^{N-1} \cos(\theta_k - \theta^*) \right) \\
    &\leq \| \theta - \theta^* \|^2, \\
    \end{split}
\end{align}
whose equality holds if and only if $\theta=\theta^*$.
Now, we can take $0<q<1$ to confirm that (\ref{Contraction mapping}) holds.

Consequently, we show that, in the neighborhood of the solution, our algorithm has an infinitesimal error $\varepsilon_n$ such that
\begin{align}
    \left\| \theta^{(n+1)} - \theta^* \right\|
    &<
    C \left\| \theta^{(n)} - \theta^* \right\|^2 + \varepsilon_n,
\end{align}
and $\varepsilon_n$ goes to zero as $\theta^{(n)}$ approaches the optimal solution.
We will examine the performance of our algorithm in Sec.~\ref{Sec:main-results} using numerical examples.

\subsection{Gauss-Newton based quantum algorithm (GNQA)}\label{Subsec:main-algorithm}

Now, we present the main algorithm of this paper.
To improve the convergence property of the Gauss-Newton method, we update the value of $\eta$ at each step.
More concretely, we introduce a variable step size $\eta$ in our iterations in a following manner,
\begin{align}
    \begin{split}
    \theta^{(n+1)}
    &= \theta^{(n)} + \eta \, J^T_{\varphi}(\theta^{(n)}) R_f ( \ket{\varphi(\theta^{(n)})} ), \label{main_algorithm}\\
    \ket{\zeta^{(n)}} 
    &= 
    R_f \left( \ket{\varphi(\theta^{(n)})} \right), \\
    \eta 
    &= \frac{1}{\braket{\varphi(\theta^{(n)}) | \zeta^{(n)}}}.
    \end{split}
\end{align}
This is the Gauss-Newton based quantum algorithm (GNQA).

Now, we briefly discuss the justification for using a variable step size in Eq.~(\ref{main_algorithm}).
To simplify, we consider a case where the optimal solution is $\ket{\xi^*} = (1, 0, \cdots, 0)$ with $\theta^* = (0, \cdots, 0)$, and $\ket{\zeta^{(n)}} = \ket{\xi^*}$ at each step.
Then, using Eq.~(\ref{main_algorithm}) we arrive at the following recurrence relation for each component of optimization parameter
\begin{align}
    x^{(n+1)} = x^{(n)} - \frac{\sin(x^{(n)})}{\cos(x^{(n)})}.
\end{align}
Note that here, we apply $\eta = \cos(\omega)^N$ to Eq.~(\ref{convergence_along_line_0}).
This recurrence relation is just the Newton's method for finding a root of the sin function, so if the initial point is set to $x_0 = \pi/4$, $x^{(n)}$ will approach zero, cubically, as $n \rightarrow \infty$. 
Here we use $x - \tan(x) = x^3 /3 + O(x^5)$.
Therefore, our algorithm, Eq.~(\ref{main_algorithm}), modifies and improves the convergence property of the standard Gauss-Newton iteration with $\eta=1$.
This argument implies that, under an optimal condition, our algorithm can find the optimal solution at single- or double-precision within 4 iterations independent of the problem size $N$.

\subsection{The Quantum Algorithm}\label{Subsec:Q-comp}

We conclude this section with a description of the quantum algorithm.
Our algorithm is formulated in the Hilbert space $\mathcal{H}$, and thus, it is a convenient footing for quantum computation.
More concretely, the circuit parameters are updated by preparing the state $\ket{\zeta^{(n)}}$ and calculating $y = J_\varphi^T(\theta^{(n)}) \ket{\zeta^{(n)}}$.
Since the $k$-th element of $y^{(n)}$ is given by:
\begin{align}
    y^{(n)}_k 
    &= \braket{\partial_k \varphi(\theta) | \zeta^{(n)}},
\end{align}
where $\ket{\partial_k \varphi(\theta)}$ is defined by
\begin{align}
    \ket{\partial_k \varphi(\theta)} 
    &= 
    \ket{s(\theta_1)} \otimes \cdots \otimes 
    \dfrac{d}{d \theta_k} \ket{s(\theta_k)}
    \otimes \cdots \otimes \ket{s(\theta_N)} \nonumber \\
    &= 
    \ket{s(\theta_1)} \otimes \cdots \otimes 
    \ket{s(\theta_k + \frac{\pi}{2})}
    \otimes \cdots \otimes \ket{s(\theta_N)},
\end{align}
for each iteration step, the necessary information to update the parameter $\theta$ can be obtained by measuring $y_k$ for $k=0, \cdots, N-1$.

Furthermore, we can use expectation values of $f(H)$ to evaluate $y$ and $\eta$ rather than inner products.
Using the orthonormal basis $\{ e_k \}_{k=0}^{N-1}$ of the parameter space $\mathcal{K}$, we have
\begin{align}
    \ket{\partial_k \varphi(\theta)} = \ket{\varphi(\theta + \frac{\pi}{2} e_k)}, 
\end{align}
resulting in 
\begin{align}
    y_k
    &= \frac{1}{C} \Big(
    \braket{\varphi_k^{+} | f(H) | \varphi_k^{+}} - \braket{\varphi_k^{-} | f(H) | \varphi_k^{-}}
    \Big), \\
    C &=
    \| f(H) \ket{\varphi(\theta)} \|^2,
\end{align}
where 
\begin{align}
    \ket{\varphi_k^{\pm}} = \ket{\varphi(\theta \pm \frac{\pi}{4} e_k)}.
\end{align}
Consequently, we have the following equivalent iteration:
\begin{align}
    \begin{split}
    \theta^{(n+1)} &= \theta^{(n)} + \eta \, y^{(n)}, \\
    \eta &= \frac{1}{\braket{\varphi(\theta^{(n)}) | f(H) | \varphi(\theta^{(n)})}}, \\
    y^{(n)}_k &= \frac{1}{2} \left[
    \braket{\varphi_k^{+} | f(H) | \varphi_k^{+}} - \braket{\varphi_k^{-} | f(H) | \varphi_k^{-}}
    \right].
    \end{split}
\end{align}
This indicates that only measurements of expectation values are required at each step. 
Note that the objective function can be efficiently calculated using a classical computer.

\section{Construction of appropriate transformations of Hamiltonian} \label{Sec:construction-f(H)}

In this section we describe how a transformation of $H$ can be constructed for the GNQA.
Once the form of $f$ has been determined, we can rely on techniques of the quantum signal processor (QSP) to prepare $f(H)$ that has an appropriate polynomial expansion formula for $f$ on a quantum computer \cite{low2019hamiltonian}.
Alternatively, the methods discussed in Ref.~\cite{subramanian2019implementing} or quantum singular value transformation \cite{gilyen2019quantum, lloyd2021hamiltonian, martyn2021grand} can be applied.
The techniques to realize $f(H)$ on a quantum system are also summarized in Ref.~\cite{lin2022lecture}.
In addition, if $f$ can be formulated with an effective Fourier expansion formula, a linear combination of unitary operators (LCU) is also a good strategy.

\subsection{Construction using monotonically decreasing functions}

First we present a simple construction of $f$ that uses monotonically decreasing positive functions to enforce Property (\ref{prop_f}).
The motivation is based on the fact that for a continuous function $L$ on a compact $\Omega \in \mathbb{R}$ and a monotonically decreasing positive function $g$, the function given by
\begin{align*}
    f(x) = C^{-1} g(L(x))^p, 
    \quad C = \int_\Omega g(L(x))^p \, dx
\end{align*}
is an appropriate nascent delta function in the sense that 
\begin{align}
    \lim_{p \rightarrow \infty} \int_\Omega L(x) f(x) \, dx = L(x^*),
\end{align}
where $x^*$ is a global minimum of $L$.

\subsubsection{Construction using power functions}

A power function is a simple way to construct $f$ that possesses the Property (\ref{prop_f}).
Specifically, for $p>0$, we define $f$ as
\begin{align}
    f(H) = \left( I - \hat{H} \right)^p,
\end{align}
where $\hat{H}$ denotes a normalization of $H$. 
Then, by defining $\hat{\lambda}_k$ as an eigenvalue of $\hat{H}$, we have
\begin{align}
    \frac{f(\lambda_k)}{f(\lambda_0)}
    =
    \left( \frac{1 - \hat{\lambda}_k}{1 - \hat{\lambda}_0} \right)^p < 1.
\end{align}
If $p$ is chosen sufficiently large, the right side approaches zero, and so does $r$. 
Therefore, $R_f(\ket{\varphi})$ becomes a good approximation of the ground state with large $p$. 
Since $\Vert H \Vert_F$ can be efficiently computed, it is reasonable to normalize $H$ such that $\hat{H} = H / \Vert H \Vert_F$. 
However, if the dimension of $M$ is large, all eigenvalues of $\hat{H}$ crowds around zero.
Therefore, $p$ must be quite large for our algorithm to work well with such a normalization. 
If we can estimate the width of $d_H$ by using its variance as discussed in Sec.~\ref{Subsec:natural-gradient}, it is more suitable to normalize $H$ using this width.

Effective quantum algorithms for constructing powers of $H$ are proposed and developed in Ref.~\cite{PRXQuantum.2.010333}.
Furthermore, the quantum algorithms using powers of the Hamiltonian for combinatorial optimization problems are also studied in Ref.~\cite{aulicino2021state}.

\subsubsection{Construction using exponential functions}

A more effective construction of $f$ is to use exponential functions. 
Specifically, for $p>0$, we can use the following function:
\begin{align}
    f(H) = e^{-p H}.
\end{align}
Then, we have
\begin{align}
    \frac{f(\lambda_k)}{f(\lambda_0)}
    =
    e^{-p(\lambda_k - \lambda_0)}.
\end{align}
Therefore, we have a good approximation of the ground state with sufficiently large $p$.
When the difference between $\lambda_0$ and the other eigenvalues becomes smaller, larger value of $p$ must be used.
In passing we note that the quantum algorithms to prepare the exponential function of Hamiltonian are also discussed in Refs.~\cite{zoufal2021error, zoufal2021variational, alghassi2021variational}. 
Finally, we can also use the LCU by discretizing the Fourier integral representation of the Laplace kernel, $f(x) = e^{-p \vert x \vert}$.

Alternatively, the Gibbs state of the Hamiltonian may be also useful to construct a function $f$ which fits our purpose:
\begin{align}
    f(H) = \frac{e^{-p H}}{\mathrm{Tr}(e^{-p H})}.
\end{align}
The Gibbs state can be prepared on a quantum computer by using the variational algorithm described in Ref.~\cite{chowdhury2020variational}.

\subsection{Construction using approximate minimal eigenvalues}

Another effective approach is to use the approximate values of lowest eigenvalues $\lambda_0$ and construct a Dirac delta function $\delta(x - \lambda_0)$.
But since it is hard to even estimate $\lambda_0$ accurately, we start our discussion by introducing an effective way to approximate $\rho$, where $\rho \approx \lambda_0$ and $\rho < \lambda_0$.

\subsubsection{Effective approximation of minimal eigenvalues}

We now discuss how $\rho$ can be approximated using a quantum computer.
We begin by defining a function $\rho$ as
\begin{align} \label{approx_min_eig}
    \rho(s) = -\frac{1}{s} \ln \left( \mathrm{Tr}(e^{-s H}) \right).
\end{align}
This approximation has the following properties beneficial to our current purpose:
\begin{align*}
    &\text{(a)} \quad 0 < s_k < s_{k+1} \Rightarrow \rho(s_{k}) < \rho(s_{k+1}) \\
    &\text{(b)} \quad \lambda_0 - \frac{1}{s} \ln(M) < \rho(s) < \lambda_0 \\
    &\text{(c)} \quad \lim_{s \rightarrow \infty} \rho(s) = \lambda_0
\end{align*}
Property (a) follows immediately from the relation,
\begin{align}
    \mathrm{Tr} ( e^{-s_k H} )
    >
    \mathrm{Tr} ( e^{-s_{k+1} H} ),
\end{align}
in combination with a property of logarithmic function.
From a relationship,
\begin{align}
    e^{-s \lambda_0} < \mathrm{Tr} (e^{-s H}),
\end{align}
we can imply that $\rho(s) < \lambda_0$.
On the other hand, since we have
\begin{align}
    \rho(s) - \lambda_0
    = -\frac{1}{s} \ln \left( \mathrm{Tr}( e^{-s(H - \lambda_0 I)} ) \right)
    < -\frac{1}{s} \ln(M)
\end{align}
we verify (b) and, consequently, (c).

Note that if $H$ is diagonal, we can compute $\rho$ from the expectation value,
\begin{align}
    \rho(s) = -\frac{1}{s} \ln \left( \braket{\varphi_0 | e^{-s H} | \varphi_0} \right) 
    - \frac{1}{s} \ln(M).
\end{align}
Moreover, by employing a Chebyshev expansion
\begin{align}
    e^{-sx} = I_0(s) + 2 \sum_{k=1}^{\infty} I_k(-s) T_k(x),
\end{align}
we can calculate the function using on the expectation values of the Chebyshev polynomials of the Hamiltonian, effectively, on a quantum computer. 
Here, $I_k$ is the modified Bessel Function of the first kind.

Another approximation algorithm for minimal eigenvalues can be found in Ref.~\cite{yalovetzky2021nisq} 
(see Algorithm 1 and 2).

\subsubsection{Construction using kernel functions}

A reasonable function to be used for $f$ is the Fourier series expression of the delta function, Dirichlet kernel, which allows us to use LCU,
\begin{align}
    f(H) 
    = \frac{1}{2 \pi} \sum_{n=-p}^{p} e^{i \pi n (H - \rho I)}
\end{align}
where $H$ is normalized such that $\vert \lambda_k \vert < 1$ for all $k$.
If $\rho$ is equal to $\lambda_0$, then $f$ approaches a delta function centered at $\lambda_0$

Alternatively, we can use the truncated Chebyshev polynomial expansion of the delta function ~\cite{childs2017quantum} to arrive at
\begin{align}
    f(H) =
    \frac{2}{\pi} 
    \left( \frac{1}{2} + \sum_{k=1}^{p} T_k(\rho) T_k(H) \right),
\end{align}
where $H$ is normalized so that $\vert \lambda_k \vert < 1$ for all $k$.

\subsubsection{Construction as resolvent operator}

One of the most effective ways to construct $f$ is to use the inverse operator.
Since a Dirac delta function can be treated as a hyper function \cite{imai1992applied} by using $1/z$, it is reasonable to use the inverse operator, namely the resolvent $1/(z - \lambda_0)$.
Since we can only use the approximate value of $\rho$, we construct an appropriate function by using $p$ so to get:
\begin{align} \label{resolvent_operator}
    f(H) = \left( H - \rho I \right)^{-p}.
\end{align}
Due to the property $\rho < \lambda_0$, the function $f(x)=1/(x-\rho)^p$ is analytic on $(-\|H\|, \|H\|)$.
Therefore, we can apply QSP or QSVT to prepare $f$ on a quantum computer.

We can also consider using the Fourier expansion for the inverse operator as described in Ref.~\cite{kyriienko2020quantum}.
More precisely, we can discretize the following integral expression induced by the combination of the properties of Hermite functions and the Gauss integration:
\begin{align}
    f(H) = 
    C \int_0^\infty \int_{-\infty}^{\infty} 
    y^{p-1} z e^{-z^2/2} e^{-iyz (H-\rho I)} \, 
    \mathrm{d}z \, \mathrm{d}y,
\end{align}
where $C$ is a constant real value.


\begin{figure}[t]
\centering
    \includegraphics[width=8.0cm]{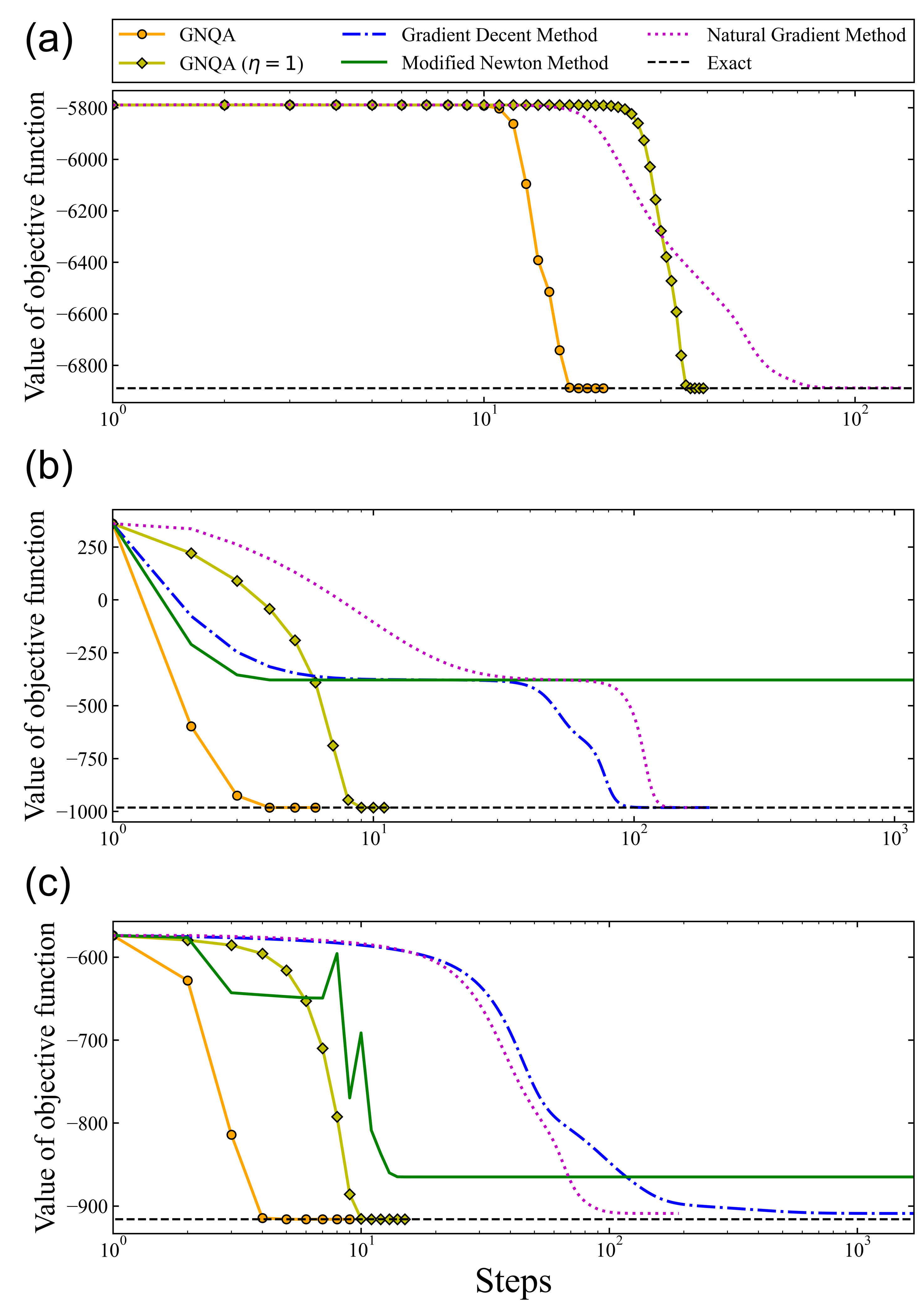}
    \caption{\label{Fig:main_convergence_1} 
    We plot the convergence behavior of the algorithms discussed in this paper, namely, GNQA, GNQA with $eta$ held constant at $1$ (GNQA ($\eta$=1)), the gradient descent method, the modified Newton's method, and the natural gradient method.
    The exact value of the optimized objective function is also plotted for reference.
    The optimization problems are (a) Number Partitioning problem, (b) Quadratic Assignment problem, and (c) General 0/1 Programming problem. 
    Note that in (a), the results of the gradient decent method (blue dashed dotted line) and the modified Newton's method (green solid line) are excluded because the value of their objective function does not change from the starting value. 
    }
\end{figure}

\begin{figure}[t]
\centering
    \includegraphics[width=8.0cm]{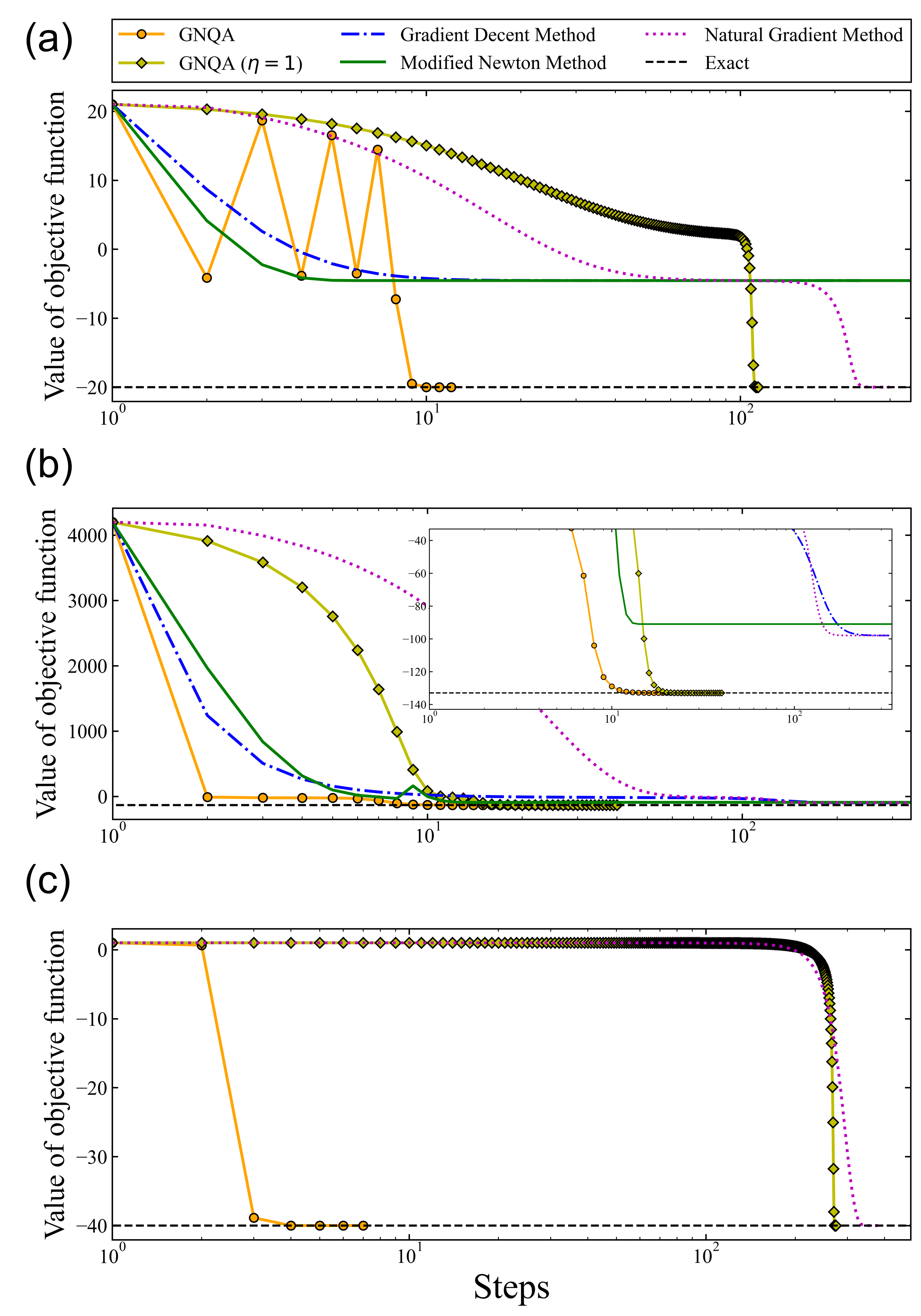}
    \caption{\label{Fig:main_convergence_2} 
    We plot the convergence behavior of the algorithms discussed in this paper for various model problems, including (a) Graph Coloring Problem, (b) General QUBO Problem, and (c) Max Cut Problem. 
    The algorithms considered are the same as those in Fig.~\ref{Fig:main_convergence_1}.
    Note that in (c), the results of the gradient decent method (blue dashed line) and the modified Newton's method (green solid line) are excluded.
    See Fig.~\ref{Fig:main_convergence_1} for the reasoning.
    }
\end{figure}

\begin{figure*}[t]
\centering
    \includegraphics[width=15.0cm]{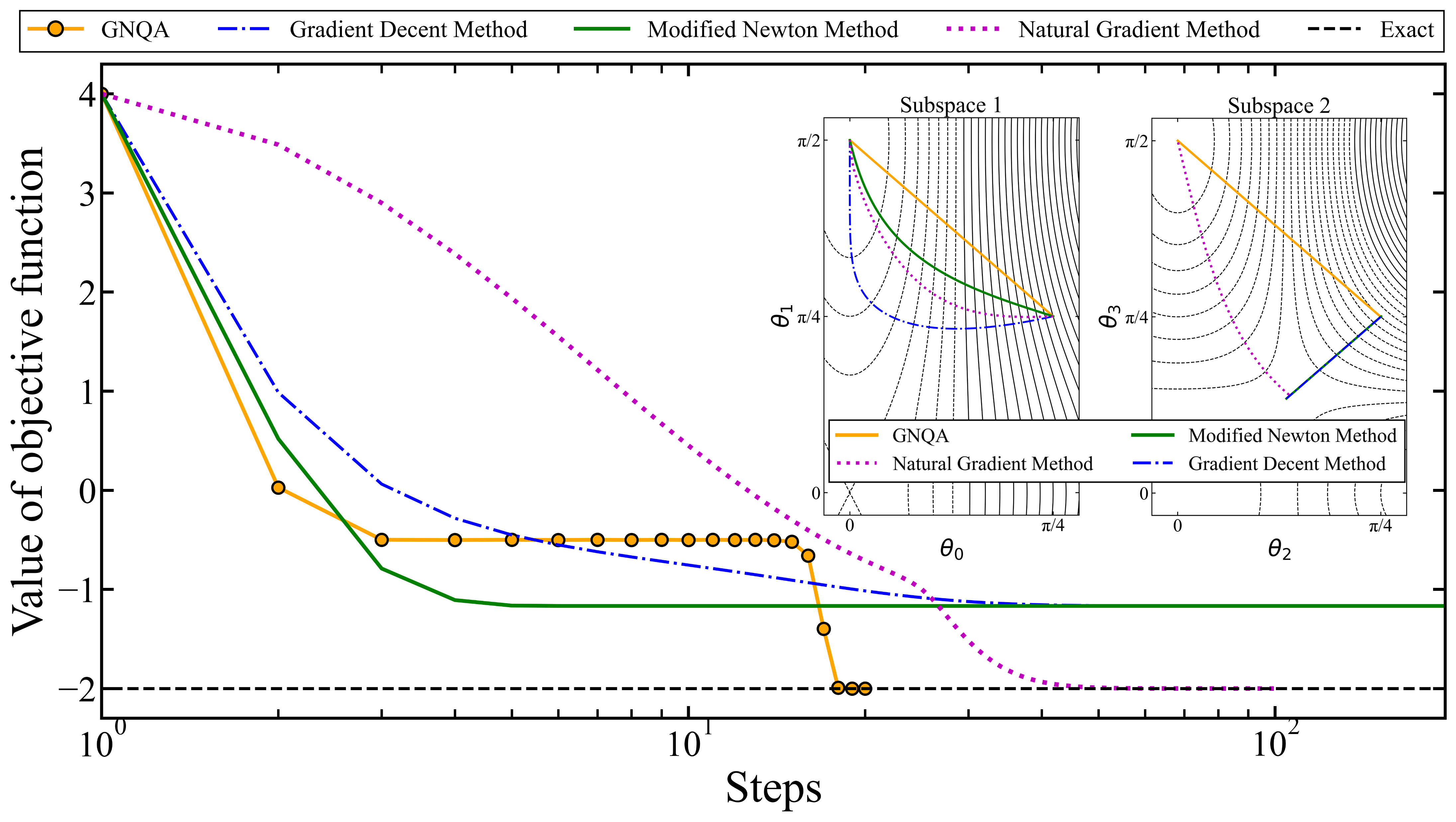}
    \caption{\label{fig:comparison_algorithms} 
    The convergence behavior of the GNQA, the gradient descent method, the modified Newton's method, and the natural gradient method are plotted for the Set Packing Problem ($N=4$).
    In the insets, we plot the 2 dimensional landscape of the objective function along two varying coordinates, namely Subspace 1 with $(\theta_0, \theta_1)$-coordinates (left) and Subspace 2 with $(\theta_2, \theta_3)$-coordinates (right).
    }
\end{figure*}

\begin{table*}[t]
  \caption{\label{tab:table1}
  Results of the numerical experiments using the algorithm described in Eq.~(\ref{main_algorithm})
  }
  \begin{ruledtabular}
  \begin{threeparttable}
  \begin{tabular}{lrrrrl}
    \multicolumn{3}{l}{Problem information} & \multicolumn{3}{l}{Experiment results}\\
    \cmidrule{1-3} \cmidrule{4-6}
    Type
    & Size \tnote{a} 
    & Number of solutions
    & Number of iterations
    & Error \tnote{b} 
    & Figure number \\
    \midrule
    QUBO                    &  4 &  1 &  2 & 2.3e-05 & \ref{fig:all}~(a) \\
    Set Packing             &  4 &  2 & 18 & 2.4e-07 & \ref{fig:all}~(b) \\
    Max-2-Sat               &  4 &  1 &  2 & 3.4e-05 & \ref{fig:all}~(c) \\
    Max Cut                 &  5 &  4 & 11 & 7.3e-04 & \ref{fig:all}~(d) \\
    Minimum Vertex Cover    &  5 &  4 & 14 & 7.1e-06 & \ref{fig:all}~(e) \\
    Set Partitioning        &  6 &  1 &  2 & 2.6e-05 & \ref{fig:all}~(f) \\
    Knapsack                &  6 &  1 &  3 & 1.5e-05 & \ref{fig:all}~(g) \\
    Number Partitioning     &  8 & 10 & 17 & 2.1e-07 & \ref{Fig:main_convergence_1}~(a), \ref{fig:all}~(h) \\
    Quadratic Assignment    &  9 &  1 &  3 & 2.0e-05 & \ref{Fig:main_convergence_1}~(b), \ref{fig:all}~(i) \\
    General 0/1 Programming & 10 &  1 &  4 & 3.3e-06 & \ref{Fig:main_convergence_1}~(c), \ref{fig:all}~(j) \\
    Graph Coloring Problem  & 15 &  6 &  9 & 5.8e-06 & \ref{Fig:main_convergence_2}~(a), \ref{fig:all}~(k) \\
    QUBO in OR-Library      & 20 &  1 & 18 & 2.9e-05 & \ref{Fig:main_convergence_2}~(b), \ref{fig:all}~(l) \\
    Max Cut                 & 25 &  4 &  3 & 3.3e-06 & \ref{Fig:main_convergence_2}~(c), \ref{fig:all}~(m) \\
  \end{tabular}
  \begin{tablenotes}
  \item[a] number of optimal parameters of the problem formulated as QUBO
  \item[b] relative error between a value of the object function that the GNQA reaches at the final step of the iteration and the exact value
  \end{tablenotes}
  \end{threeparttable}
  \end{ruledtabular}
\end{table*}

\begin{table*}
  \caption{\label{tab:table3}
  The performance of the GNQA and the QAOA on the 18-variable 2-SAT problems
  }
  \begin{ruledtabular}
  \begin{threeparttable}
  \begin{tabular}{lrrrrrr}
    \multicolumn{3}{l}{Problem information}   
    & \multicolumn{2}{l}{GNQA}  
    & \multicolumn{2}{l}{QAOA \tnote{b}} \\
    \cmidrule{1-3} \cmidrule{4-5} \cmidrule{6-7}
    & & & & 
    & \multicolumn{2}{l}{Success probability (\%) \tnote{c}} \\
    Label & Size ($N$) & Number of solutions
    & Number of iterations & Error \tnote{a} 
    & $\qquad p=1 \tnote{d}$ & $p=5$ \\ 
    \midrule
    1 & 18 & 1 & 2 & 4.0e-07 & 0.22 & 0.87 \\
    2 & 18 & 1 & 2 & 1.8e-06 & 0.33 & 1.82 \\
    3 & 18 & 1 & 2 & 2.7e-06 & 0.25 & 0.22 \\
    4 & 18 & 1 & 2 & 8.6e-06 & 0.34 & 4.15 \\
    5 & 18 & 1 & 2 & 7.0e-07 & 0.31 & 0.83 \\
  \end{tabular}
  \begin{tablenotes}
  \item[a] relative error between a value of the energy that the GNQA reach at the final step of the iteration and the exact value
  \item[b] transcribed from results reported in Ref.~\cite{willsch2020benchmarking}
  \item[c] probability of finding the ground state
  \item[d] parameter for discrete time steps of quantum annealing used in QAOA
  \end{tablenotes}
  \end{threeparttable}
  \end{ruledtabular}
\end{table*}

\section{Numerical Experiments}\label{Sec:main-results}

We have carried out numerical experiments to examine the performance of our approach described in Eq.~(\ref{main_algorithm}), and we present the results in this section.
In our experiments, we employed Eq.~(\ref{resolvent_operator}) with $p$ is fixed at $8$ to transform the Hamiltonian. 
We also approximate the minimum eigenvalue $\rho$ using Eq.~(\ref{approx_min_eig}) with a relative error of $0.1$, $\vert (\rho - \lambda_0) / \lambda_0 \vert \approx 0.1$.
Note that the initial values were set as $\theta_{\text{ini}} = (\pi/4, \ldots, \pi/4)$.
Finally, to examine the full potential of each algorithm, all the numerical experiments were carried out using the simulator. 

\subsection{Numerical experiments on representative combinatorial problems}\label{Subsec:results_1}

We first discuss the performance of GNQA on representative combinatorial optimization problems found in Ref.~\cite{glover2019quantum} and Ref.~\cite{beasley1990or}. 
The characteristics of the problems and the results of the experiments are summarized in Table \ref{tab:table1}.
In all experiments, GNQA converges to the exact solution (or one of the true optimal solutions).

Next, we compare the performance of GNQA against different algorithms, namely, VQE based on the gradient descent method (Sec.~\ref{Subsec:gradient-descent}) and the modified Newton's method (Sec.~\ref{Subsec:modified-newton}), and the generalized natural gradient method described in Sec.~\ref{Subsec:natural-gradient}.
Note that for these calculations, the step size was fixed at $\eta=1$ and $\eta=0.1$ for the modified Newton's method and the gradient descent method, respectively.
To make a fair comparison, the initial value of $\theta_{\text{ini}} = (\pi/4, \ldots, \pi/4)$ was used for all calculations.
The convergence behavior of various algorithms for some of the representative problems discussed in Table~\ref{tab:table1} is shown in Figs.~\ref{Fig:main_convergence_1} and \ref{Fig:main_convergence_2}. 

In Figs.~\ref{Fig:main_convergence_1}(a-c), we plot the results for number partitioning problem, quadratic assignment problem, and General 0/1 Programming, having the problem sizes, 8, 9, and 10, respectively.
Note that in Fig.~\ref{Fig:main_convergence_1}(a), we excluded the results of the gradient decent method (dot-dashed blue curve) and the modified Newton's method (solid green curve) because their objective function did not change from the initial value.
The modified Newton's method gives the poorest performance since it is not able to find the optimal solution in any of the three problems.
The generalized natural gradient method does reasonably well for all problems, but it requires more iterations than the GNQA.
Clearly, our approach, the GNQA, outperforms other methods, finding the optimal solution in the smallest number of iterations.

The convergence behavior for larger problems are shown in Fig.~\ref{Fig:main_convergence_2}.
These results are for (a) graph coloring problem, (b) QUBO in OR-Library, and (c) max cut problem, having the problems sizes, 15, 20, and 25, respectively.
We excluded the result of the gradient descent method from Fig.~\ref{Fig:main_convergence_2}(c) because its objective function value did not change from the initial value.
Once again, the gradient descent method and the modified Newton's method give poor performance, failing to find the optimal solution in all cases.
The natural gradient approach gives a reasonable performance, but our approach, particularly the GNQA, gives the best performance, finding the optimal solution in fewest number of iterations.
Note that the GNQA shows an oscillatory behavior in Fig.~\ref{Fig:main_convergence_2}(a).
This occurs when the ground state is degenerate; as the trajectory feels the pull from multiple solutions, the algorithm is deciding which solution to pursue.

We examined the convergence behavior for all the problems listed in Table~\ref{tab:table1}, and the results are discussed in the Appendix (See Fig.~\ref{fig:all} in the Appendix~\ref{Appendix:all-results}).
The results demonstrate that the GNQA outperforms all other methods in both convergence properties and accuracy for all the problems considered here.

\subsection{Visualized comparison among different methods of solving combinatorial problems} \label{Subsec:results_2}

Now, we highlight the characteristics of the GNQA.
In Fig.~\ref{fig:comparison_algorithms} we present the results of our numerical simulations on the Set Packing problem.
Here, the result of GNQA is compared to those obtained using the VQE based on both the gradient descent method and the modified Newton's method, and the generalized natural gradient method.
We plot the value of the objective function as a function of steps taken in the optimization.
Furthermore, in the insets of Fig.~\ref{fig:comparison_algorithms}, we plot the paths taken by various algorithms on a contour map, and note that here we plot the trajectories along two varying parameter spaces, $(\theta_0, \theta_1)$-coordinates and $(\theta_2, \theta_3)$-coordinates.

It is clear from the inset of Fig.~\ref{fig:comparison_algorithms}, Subspace 1, that the trajectory of the gradient descent method (dot-dashed blue curve) passes through a plateau.
On the other hand, the modified Newton's method successfully avoids this plateau.
However, for both these methods, the gradient of the problem becomes zero (see Sec.~\ref{Subsec:modified-newton} for more detail), which implies that there is a risk of the solution converging to a stationary point. This is illustrated in the other inset of Fig.~\ref{fig:comparison_algorithms}, Subspace 2, where both the gradient descent and the modified Newton's method find a stationary point rather than the optimal solution.

Finally, for this problem, the generalized natural gradient method (dotted magenta curve) converged to the optimal solution without being trapped in a stationary point. However, it is important to note that the method still has a risk of finding a local minimum rather than the optimal solution when the method is applied to problems of larger size (see Fig~\ref{Fig:main_convergence_1}(c))

While other methodologies must overcome the problems associated with local minima or plateau, the GNQA does not suffer from these issues.
This is demonstrated by insets of Fig.~\ref{fig:comparison_algorithms}, where we observe that the GNQA converges immediately to the solution, taking a straight path to the optimal solution from the initial point.
Note that the ground state for this problem is degenerate, and when there are multiple optimal solution, the trajectory feels the pull from these solutions.
This is reflected in the convergence behavior of the GNQA algorithm shown in Fig.~\ref{fig:comparison_algorithms} where there is a small oscillation in the value of the objective function before it finds the optimal solution.
In this case, there are two optimal solutions that differ along the $(\theta_2, \theta_3)$-coordinates, subspace 2 in Fig.~\ref{fig:comparison_algorithms}, namely the points $(0, \pi/2)$ and $(\pi/2, 0)$.
For this reason, the value of the objective function oscillates while the trajectory finds and picks one solution to pursue.

\subsection{Comparison of GNQA with QAOA}\label{Subsec:results_3}

We have compared the performance of our quantum algorithm with QAOA using the combinatorial optimization problems given in Ref.~\cite{willsch2020benchmarking}, for which performance of QAOA has been measured.
It is reported in the paper that they employed the Nelder–Mead algorithm \cite{nelder1965simplex} for an effective optimizer of QAOA. 
For our study, we picked 5 problems out of the 2-SAT problems each having 18 variables, and these problems had the largest number of variables among the problems discussed in the paper.
These problems have one optimal solution with the minimum energy of $-19$ (see Ref.~\cite{willsch2020benchmarking} for the details).
Note that we have decided to carry out our calculation on 5 out of 6 problems because the data presented in Ref.~\cite{willsch2020benchmarking} seems to be incorrectly transcribed, having multiple optimal solutions with the energy other than $-19$.
For another extensive study on the performance of the QAOA see Ref.~\cite{zhou2020quantum}.

We present our result in Table \ref{tab:table3} along with those of the QAOA found in Ref.~\cite{willsch2020benchmarking}. 
These results show that our algorithm gives optimal solutions with sufficient accuracy after only two iterations.
Furthermore, our algorithm is able to find the optimal solution in all of the cases considered here.
Meanwhile, QAOA does not enjoy such a success in finding the true ground state.
Therefore, we conclude that our algorithm gives a clear advantage over QAOA in solving combinatorial optimization problems.

\section{Discussions}\label{Sec:discussion}

The potential application of our algorithm is not limited to QUBO problems, and here we briefly discuss other potentials of our approach.

\begin{figure}[t]
\centering
    \includegraphics[width=7.5cm]{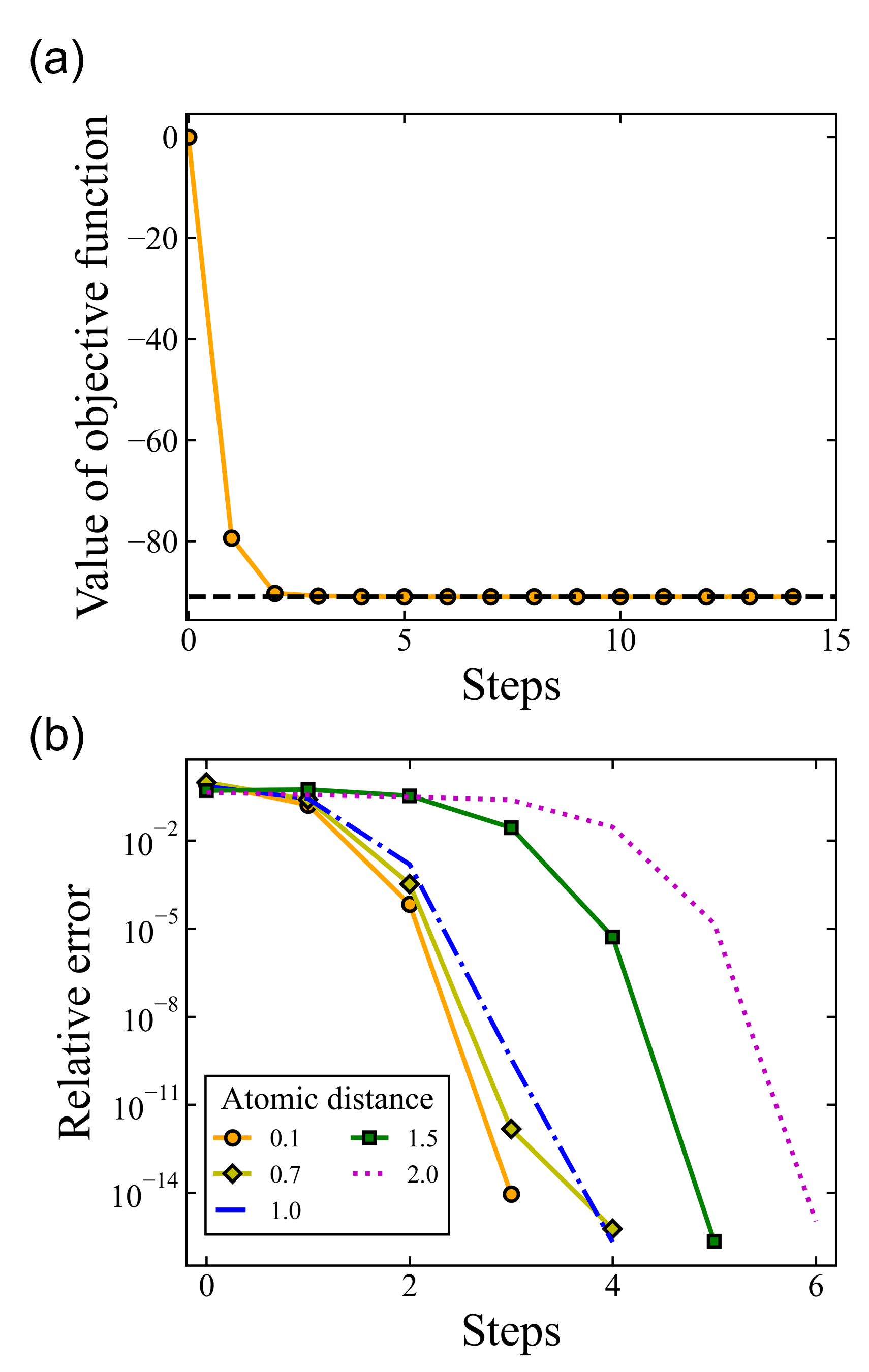}
    \caption{\label{Fig:discussion}
    (a) The convergence behavior of the GNQA for the 3-SAT problem, having 20 variables, 91 clauses, and 8 optimal solutions, is plotted.
    The black dashed line represents the optimal value of 0.
    (b) The convergence behavior of the GNQA is plotted for the electronic structure problem of hydrogen molecule with different internuclear distances (\AA).
    }
\end{figure}

\subsection{Higher order unconstrained binary optimization problems}

Our algorithm can be extended to higher order unconstrained binary optimization problems, also known as polynomial unconstrained binary optimisation, in a straightforward manner.
The problem can be formulated by the Hamiltonian of the form,
\begin{align}
    H = \sum_i c_i \sigma^z_i + \sum_{ij} c_{ij} \sigma^z_i \sigma^z_j
    + \sum_{ijk} c_{ijk} \sigma^z_i \sigma^z_j \sigma^z_k + \cdots.
\end{align}
Note that any higher order problems can be reduced to a quadratic one, but one needs to increase the number of parameters inorder to be reformulated as a QUBO problem \cite{biamonte2008nonperturbative}.
Therefore, our algorithm provides an advantage here.
We have confirmed that our algorithm is able to solve the 3-SAT problem, having 20 variables, 91 constraints, and 8 optimal solutions, and the results are shown in Fig.~\ref{Fig:discussion}(a).
Since there are many important and relevant optimization problems in the real world that require higher order terms to describe various complex interactions \cite{babbush2012construction, o2015bayesian, xia2017electronic, robert2021resource}, we argue that our approach in an effective way to solve a wide range of the combinatorial optimization problems.

\subsection{Electronic structure problems in quantum chemistry}

Another potential application is in the field of quantum chemistry \cite{kandala2017hardware}.
Our approach can be reformulated to variationally solve the electronic structure problem for the Hydrogen molecule, \ce{H2}, with some modification.
More concretely, we can permute several components of the state vector given in Eq.~(\ref{ansatz}), and we can construct an ansatz that can express the ground state of \ce{H2}.
Since the Riemannian metric (= $J^T J$) is invariant under such a permutation, our algorithm can be applied without any modifications.
The convergence behavior of our algorithm for the electronic structure problem of \ce{H2} at varying internulcear distances are plotted in Fig.~\ref{Fig:discussion}(b).
Similar to our previous observations for the combinatorial optimization problems, the results plotted show that our algorithm converges rapidly to the optimal solution for all internulcear distances.

\section{Conclusion}

We developed a new algorithm for quantum optimization problems that can efficiently solve QUBO problems.
The value and the effectiveness of our method is demonstrated by the numerical experiments we carried out, where we confirmed the advantage of our algorithm over other methods that were investigated in this paper.
The results show that the quantum algorithm proposed in this paper converges to one of the optimal solutions rapidly without being trapped in local minima or plateaus.

\begin{acknowledgments}
The authors would like to thank Noriaki Shimada for constructive discussions regarding applications of the method to various fields of optimization problems.
The authors further acknowledge the generous support from Yasuyuki Nishibayashi.

IBM, the IBM logo, and ibm.com are trademarks of International Business Machines Corp., registered in many jurisdictions worldwide. Other product and service names might be trademarks of IBM or other companies. The current list of IBM trademarks is available at https://www.ibm.com/legal/copytrade.
\end{acknowledgments}

\bibliography{main}

\appendix

\section{Geometric aspect of Newton Methods} \label{Appendix:Geo-Newton}

We describe some essential properties of dynamics which are effective for the analysis of non-linear equations.
Let $F \colon \mathbb{R}^N \rightarrow \mathbb{R}^N$ be a smooth function and $J$ be the Jacobian with variable $w \in \mathbb{R}^N$, that is, $J = \partial F / \partial w$.
To simplify the discussion, we only deal with the well-posed case that there exists a connected closed subset $\Omega \subset \mathbb{R}^N$, where $J$ is full-rank and the equation has a unique solution $\xi$.
Therefore, the positive matrix $G = J^T J$ induces a Riemannian metric $g$ on $\Omega$ and $(\Omega, g)$ becomes a Riemannian manifold under some appropriate conditions.
A Newton method in the framework of Riemannian geometry is also studied in \cite{adler2002newton}.

Let $L$ be a Lagrangian given by 
\begin{equation}
	L \left( w, v \right) = \frac{1}{2} \, v^T G(w) v
\end{equation}
with $v = dw/dt$ (also written as $\dot{w}$).
The Euler-Lagrange equation for $L$ is then expressed as
\begin{equation} \label{Euler_Lagrange_EQ}
	\frac{d p}{d t} - \nabla_w L = 0
\end{equation}
with momentum vector $p = G v$.
If the boundary condition at two points in $\Omega$, 
$w(t_0) = w_0, w(t_1) = w_1$,  
is imposed on (\ref{Euler_Lagrange_EQ}), 
a geodesic between $w_0$ and $w_1$ is obtained as the solution.
In contrast, if we give an appropriate initial condition, 
$w$ describes the motion along the geodesic from $w_0$ to $w_1$.
In fact, the following statement holds;

\begin{proposition} \label{Thm_GeodesicProperty}
	Let $w_0$ be an arbitrary point in $\Omega$ and 
	$(w(t), p(t))$ be a solution of equation (\ref{Euler_Lagrange_EQ})
	with the following initial condition; 
	\begin{equation}
		w(0) = w_0, \quad p(0) = - J(w_0)^T F(w_0).
	\end{equation}
	Then $w$ satisfies
	\begin{equation}
		F \big( w(t) \big) = (1 - t) F \big( w_0 \big),
	\end{equation}
	for $t \in [0,1]$.
	In particular, $w(t)$ passes through the point which is
	a solution of non-linear equation $F(w) = 0$ at $t = 1$,
	that is, $\xi = w(1)$.
\end{proposition}
We briefly describe the outline of the proof for the statement above.
Throughout this paper, we regard functions of $w$ as those of $t$ in the trivial way; for instance, $F(t)=F( w(t) )$.
Note that $p$ can be expressed as $p = J^T dF/dt$.
Then using the Beltrami identity for (\ref{Euler_Lagrange_EQ}) with the initial condition above leads to the equation
\begin{equation} \label{F_Geodesic_flow_1}
	\frac{d}{dt} F = - F_0,
\end{equation}
where $F_0 = F(0)$.
Thus, a closed form expression is obtained as
\begin{equation} \label{F_Geodesic_flow_2}
	F(t) = (1 - t) F_0,
\end{equation}
which gives $F(1) = 0$ as asserted in Theorem \ref{Thm_GeodesicProperty}.

Now we take a different expression 
that the coefficient $(1-t)$ in (\ref{F_Geodesic_flow_2}) is replaced 
by a different monotonically decreasing function, that is, 
\begin{equation} \label{F_general}
	F(t) = \rho(t) F_0,
\end{equation}
where $\rho$ denotes a monotonically decreasing smooth function 
from $(t_0, t_1)$ onto $(0, 1)$. 
Then, we give the following differential equation 
whose solution is of the closed form (\ref{F_general});
\begin{equation} \label{DEQ_F_general}
	\frac{d}{dt} F = \chi \, F, \quad
	\chi(t) = \frac{d}{dt} \ln(\rho(t)).
\end{equation}
A motion described by this differential equation differs from
the one that is described by (\ref{Euler_Lagrange_EQ}),
but these two motions are along the same geodesic.
Consequently, the differential equation
\begin{equation} \label{DEQ_general}
	J \frac{dw}{dt} = \chi \, F, \quad t \in (t_0, t_1)
\end{equation}
with an initial condition $w_0 = w(t_0)$ has a unique solution that satisfies $F(w(t_1)) = 0$.
Furthermore, the orbit under flow $f$ defined by $f(w_0, t) = w(t)$ coincides with that of the geodesic equation (\ref{Euler_Lagrange_EQ}).
Note that since equation (\ref{DEQ_general}) is equivalent to equation (\ref{DEQ_F_general}), the orbit is invariant under coordinate transformations.

With respect to the choice of $\rho$, the end point $t_1$ can be set as $\infty$ under some appropriate conditions and the property above still holds in the sense that 
\begin{equation}
	\lim_{t \rightarrow \infty} F( w(t) ) = 0.
\end{equation}
In particular, if we set $\rho (t) = e^{-t}$, then $\chi (t) = -1$ and $F$ can be represented as $F(t) = e^{-t} F_0$.
Then, if we apply the Euler method to the differential equation (\ref{DEQ_general}) with step size $\triangle t$ in this case, the corresponding iteration step can be written as 
\begin{equation}
	w_{i+1} = w_i - \triangle t \, J(w_i)^{-1} F(w_i),
\end{equation}
which recovers the Newton method with step size $\triangle t = 1$.

\section{Results of the numerical experiments}\label{Appendix:all-results}

\begin{figure*}[t]
\centering
    \includegraphics[width=17.0cm]{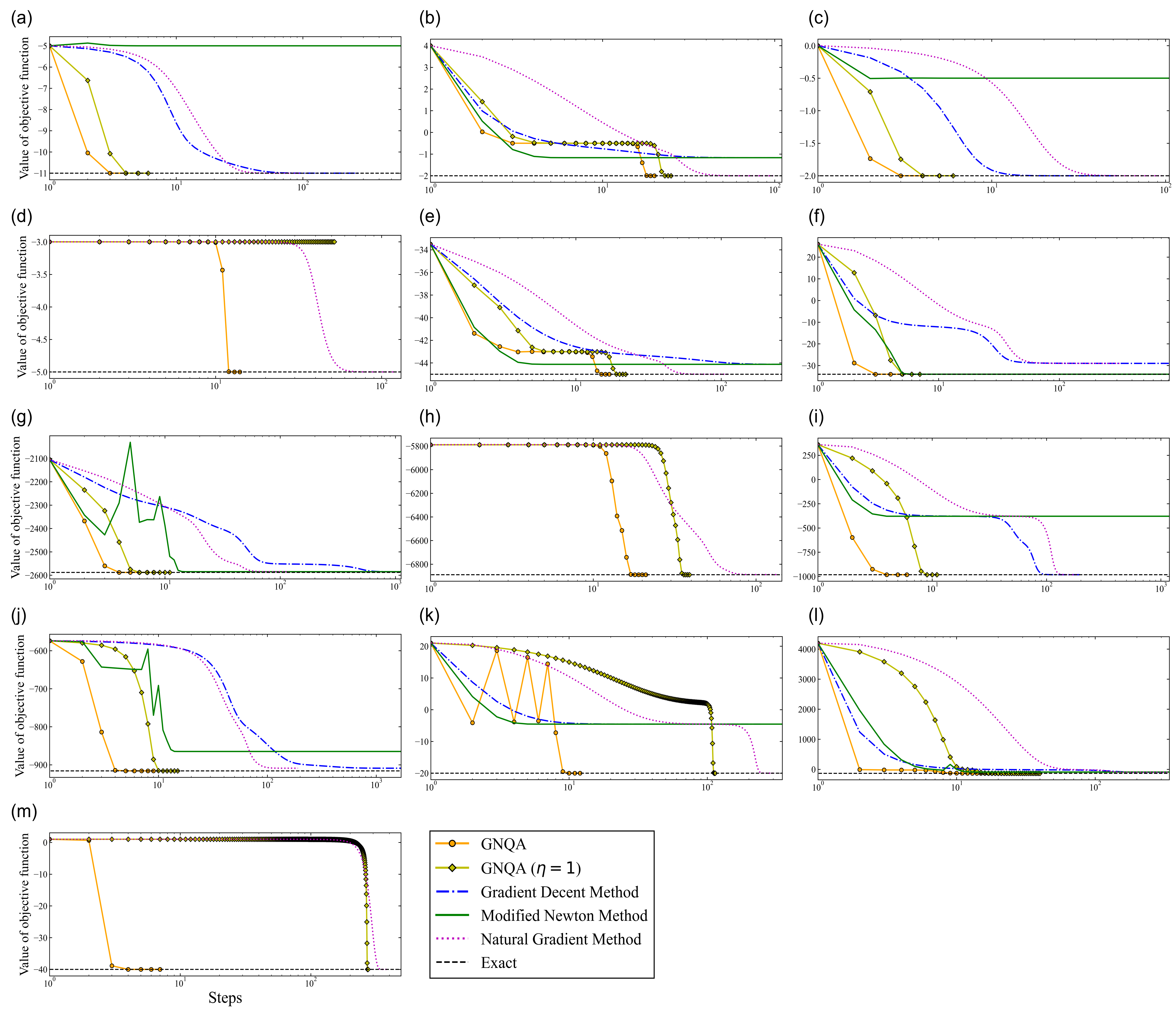}
    \caption{\label{fig:all} 
    We plot the convergence behavior of the algorithms discussed in this paper, namely, GNQA, GNQA with $\eta$ held constant at $1$ (GNQA ($\eta$=1)), the gradient descent method, the modified Newton's method, and the natural gradient method.
    The optimal value of the objective function is also plotted.
    For the assignments of the figures, see the optimization problems in Table~\ref{tab:table1}.
    }
\end{figure*}

In Fig.~\ref{fig:all}, we plot the convergence behavior of the algorithms discussed in this paper for all the optimization problems considered in Table~\ref{tab:table1}.
In addition, the accuracy is shown in Table~\ref{tab:table5}.
Note that the results of the gradient descent method (blue dashed dotted line) and the modified Newton's method (green solid line) are excluded from Figs.~\ref{fig:all}(d), (h), and (m) because the value of the objective function for those calculations did not change from the initial value.

\begin{table*}
  \caption{\label{tab:table5}
  The accuracy of the numerical experiments carried out using the algorithms discussed in this paper.
  For the details of the types problems investigated, see Table~\ref{tab:table1}.
  }
  \begin{ruledtabular}
  \begin{threeparttable}
  \begin{tabular}{lrrrrr}
    \multicolumn{1}{l}{Problems} & \multicolumn{3}{l}{Experimental results: accuracy (\%) \tnote{a}} \\
    \cmidrule{1-1} \cmidrule{2-6} 
    Label in Fig.~\ref{fig:all}
    & GNQA
    & GNQA ($\eta=1$)
    & Gradient Descent
    & Modified Newton
    & Generalized Natural Gradient \\
    \midrule
    (a)     & 100.0 & 100.0 & 100.0 &  45.5 & 100.0 \\
    (b)     & 100.0 & 100.0 &  58.3 &  58.3 & 100.0 \\
    (c)     & 100.0 & 100.0 & 100.0 &  25.0 & 100.0 \\
    (d)     & 100.0 &  60.0 &  60.0 &  60.0 & 100.0 \\
    (e)     & 100.0 & 100.0 &  98.1 &  98.1 & 100.0 \\
    (f)     & 100.0 & 100.0 &  85.3 & 100.0 &  85.3 \\
    (g)     & 100.0 & 100.0 &  99.9 &  99.9 &  99.9 \\
    (h)     & 100.0 & 100.0 &  84.0 &  84.0 & 100.0 \\
    (i)     & 100.0 & 100.0 & 100.0 &  38.6 & 100.0 \\
    (j)     & 100.0 & 100.0 &  99.2 &  94.4 &  99.2 \\
    (k)     & 100.0 & 100.0 &  22.7 &  22.7 & 100.0 \\
    (l)     & 100.0 & 100.0 &  73.7 &  68.4 &  73.7 \\
    (m)     & 100.0 & 100.0 &   2.5 &   2.5 & 100.0
  \end{tabular}
  \begin{tablenotes}
  \item[a] accuracy of a value that each method reaches at the final step of the iteration
  \end{tablenotes}
  \end{threeparttable}
  \end{ruledtabular}
\end{table*}

\section{Features of the transformation $f$}\label{Appendix:other-results}

\begin{figure}[t]
\centering
    \includegraphics[width=7.5cm]{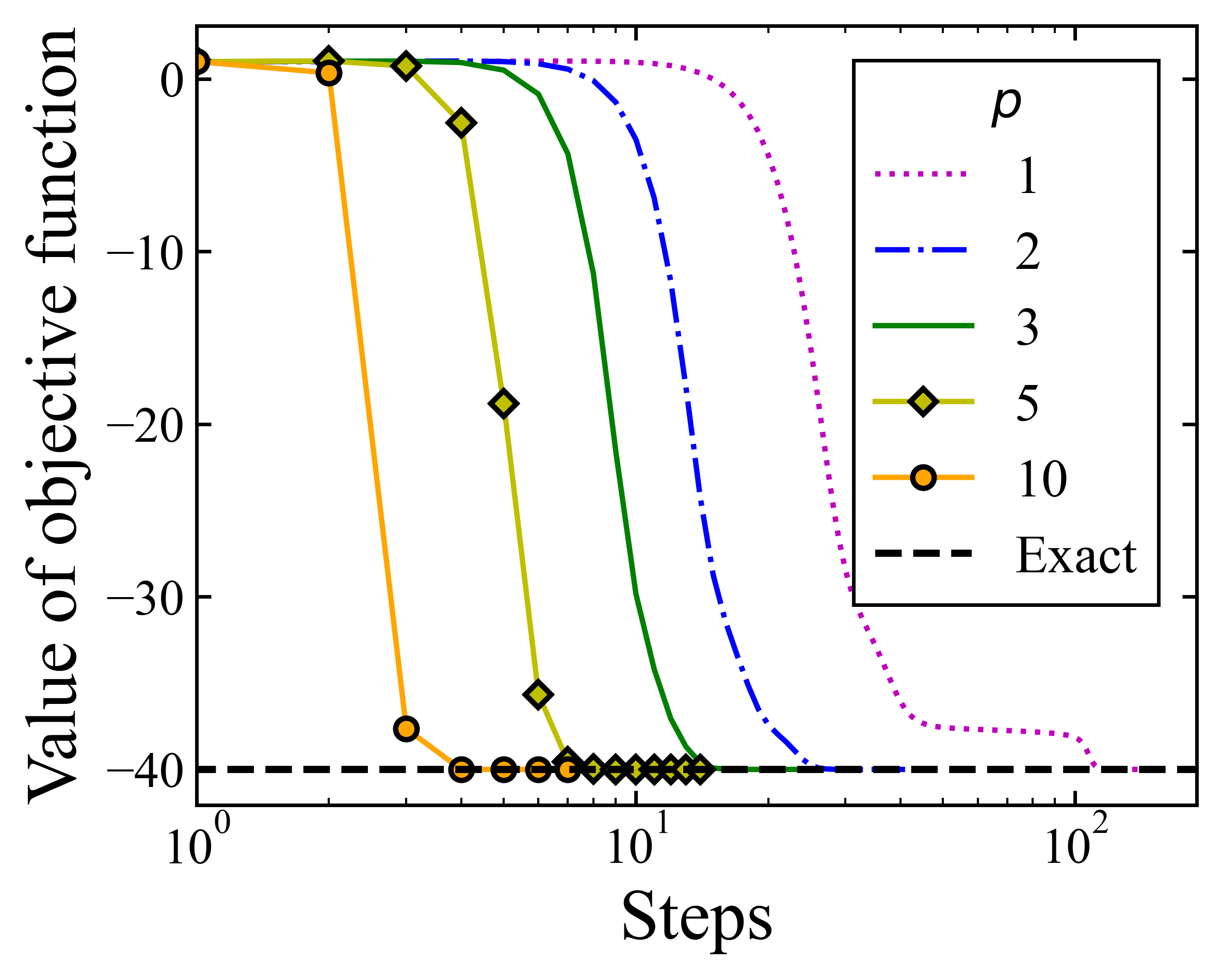}
    \caption{\label{Fig:different_p} 
    We plot the convergence behavior the GNQA with several different values of $p$.
    The problem considered here is the Max Cut problem of size $N=25$ (see Table~\ref{tab:table1}).
    }
\end{figure}

In this section, we investigate the features of the transformation defined by Eq.~(\ref{resolvent_operator}) using numerical simulations.
One interesting experiment is to see how the results are influenced by the parameter $p$, and we plot the result of such an experiment on the Max Cut problem of size $N=25$ (see Table~\ref{tab:table1}) in Fig.~\ref{Fig:different_p}.
Note that here we vary $p$ while $\sigma$ is held fixed.

\begin{figure*}[t]
\centering
    \includegraphics[width=14cm]{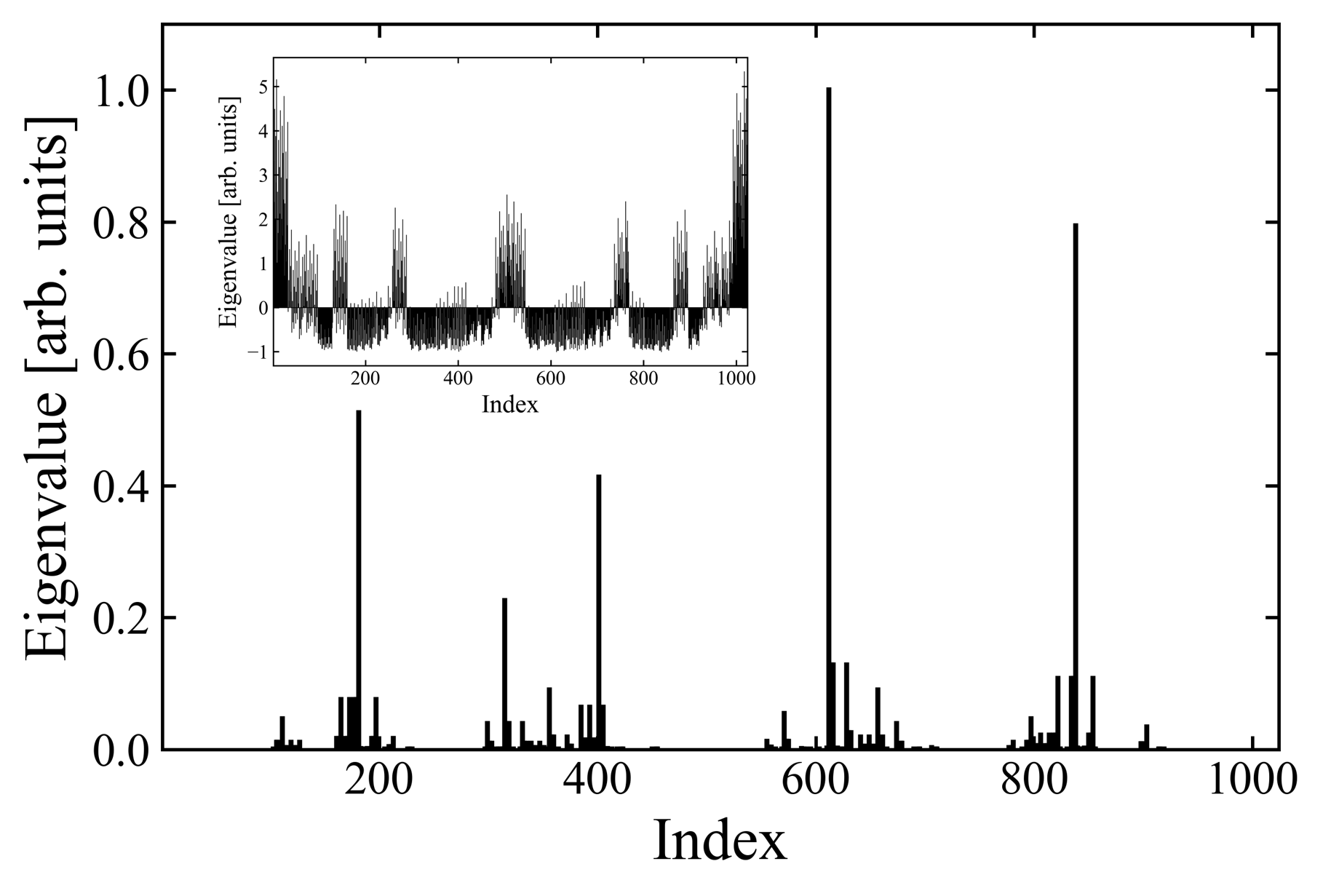}
    \caption{\label{Fig:change-eigen} 
    The distribution of eigenvalues of after a transformation with $f(H)$ is plotted. 
    The value of the eigenvalue is plotted as a function of its corresponding index.
    For comparison, the distribution of eigenvalues for the Hamiltonian $H$ is plotted in the inset.
    }
\end{figure*}

Next, in Fig.~\ref{Fig:change-eigen} we show how the distribution of eigenvalues are influenced by this transformation $f$.
Note that an ideal distribution of eigenvalues will like the discrete unit sample function.

\end{document}